\documentclass[12pt]{article}
\usepackage{physics}
\usepackage{epsf}
\usepackage{aas_macros}
\usepackage{amssymb}
\usepackage{comment}
\usepackage{amsmath}
\usepackage{braket}
\usepackage{mathrsfs}
\usepackage{mathtools}
\usepackage{array}
\usepackage{fancyhdr}
\usepackage[dvipdfmx]{graphicx}
\usepackage{color}
\usepackage{cite}
\usepackage{bm}
\usepackage{url}
\usepackage{fancyhdr}
\usepackage{here}
\usepackage{xcolor}
\usepackage{appendix}
\usepackage[colorlinks,citecolor=blue]{hyperref}
\usepackage[hang,small,bf]{caption}
\usepackage[subrefformat=parens]{subcaption}
\renewcommand{\thefootnote}{\#\arabic{footnote}}

\renewcommand{\thefootnote}{\fnsymbol{footnote}}
\def\thefootnote{\fnsymbol{footnote}}

\makeatletter

\@addtoreset{equation}{section}
\makeatother

\def\be{\begin{equation}}
\def\ee{\end{equation}}
\def\ben{\begin{eqnarray}}
\def\een{\end{eqnarray}}

\usepackage{scalerel}
\usepackage[usestackEOL]{stackengine}

\def\dashint{\,\ThisStyle{\ensurestackMath{%
            \stackinset{c}{.2\LMpt}{c}{.5\LMpt}{\SavedStyle-}{\SavedStyle\phantom{\int}}}%
        \setbox0=\hbox{$\SavedStyle\int\,$}\kern-\wd0}\int}

\begin{document}


\begin{center}

\vskip .75in

{\Large \bf Towards model-independent identification of lensed gravitational waves using Kramers-Kronig relation}

\vskip .75in

{\large
So Tanaka$\,^1$, Gopalkrishna Prabhu$\,^2$,\\
\vspace{1mm}
Shasvath J. Kapadia$\,^2$, and Teruaki Suyama$\,^1$
}

\vskip 0.25in

{\em
$^{1}$Department of Physics, Institute of Science Tokyo, 2-12-1 Ookayama, Meguro-ku,
Tokyo 152-8551, Japan\\
$^{2}$Inter-University Centre for Astronomy and Astrophysics, Post Bag 4, Ganeshkhind, Pune 411007, India
}

\end{center}
\vskip .5in

\begin{abstract}
Observations of microlensed gravitational waves (GWs) emanated by compact binary coalescences (CBCs) are essential for studying the mass density distribution in the universe, including black holes and dark matter halos. However, no confident detection of microlensed GWs have been reported to date. There are two important challenges in the identification of microlensed GWs. The first is that the source waveform and lens structure models are not known a-priori. The second is that certain classes of unlensed GWs could mimic microlensed GWs, resulting in undesirable false alarms. In this work, we propose to use the Kramers-Kronig relation for gravitational lensing systems. We argue that such systems are essentially linear response systems obeying causality, where KK relation must hold. The power of this method lies in the fact that microlensed GWs, regardless of the lens structure, must obey KK relation, while unlensed GW events are not in general expected to obey it. This, in principle, allows us to identify microlensed GWs while dismissing microlensing mimickers. We provide the first important steps towards a methodology that exploits KK relation, and test its usefulness under idealized conditions.
\end{abstract}

\renewcommand{\thepage}{\arabic{page}}
\setcounter{page}{1}
\renewcommand{\thefootnote}{\#\arabic{footnote}}
\setcounter{footnote}{0}

\section{Introduction}
The phenomenon of light or Gravitational Waves (GWs) being distorted by the gravity of a massive object is known as Gravitational Lensing (GL)\cite{Bartelmann:2010fz,schneider1999gravitational, misner2017gravitation,Bailes:2021tot}. The amplification factor, which describes the distortion of the waveform is observationally important because it contains information such as the mass density distribution of the lensing object\cite{Tambalo:2022wlm, Savastano:2023spl,Oguri:2020ldf,Takahashi:2005ug}. Although all of the GL phenomena observed to date are of light, the GL of GWs is rather crucial for the purpose of revealing the physical properties of lensing objects.
The main reason for this is that the propagation of light and GWs are described by theories in different regimes. Propagation of light with short wavelengths is based on Geometric Optics (GO), which yields only a limited number of parameters, such as magnification and time delay. On the other hand, the propagation of GWs with longer wavelengths is based on Wave Optics (WO), and frequency dependence appears in the amplification factor due to diffraction and interference\cite{Nakamura:1997sw, Nakamura:1999uwi,Takahashi:2003ix, ruffa1999gravitational}. By analyzing this frequency dependence, information on the mass density distribution of lensing objects can be obtained.
Another reason why GWs are important for GL observations is their coherence (phase alignment). Light emitted by celestial objects is a superposition of random phases, whereas GWs emitted from GW sources such as binary stars are in phase\cite{maggiore2007gravitational}. This makes it possible to directly observe the waveform of GWs, which is suitable for detailed measurement of the frequency dependence of the amplification factors.

Most studies to date rely on an approach that employs templates for both the unlensed waveform and the amplification factor, using standard source models and simple lens models\cite{Takahashi:2003ix}. Such model-dependent templated searches have been carried on LIGO-Virgo-KAGRA \cite{LIGOScientific:2014pky, TheVirgo:2014hva, KAGRA:2020tym} data produced during its first three observing runs\cite{LIGOScientific:2018mvr,LIGOScientific:2020ibl,LIGOScientific:2021usb,KAGRA:2021vkt}, although no confident detections have been reported \footnote{Typical model-dependent searches for signatures of microlensing involve large-scale Bayesian parameter estimation, although more rapid searches involving banks of templates have also been attempted \cite{Chan:2024qmb}.} \cite{LIGOScientific:2021izm, LIGOScientific:2023bwz}. In this framework, the detection of GL signals and the measurement of the amplification factor are achieved by identifying the source and lens parameters that best match the observed waveform. The amplification factor derived in this manner is physically meaningful, as it corresponds to a physical lens model. In contrast, our study adopts an alternative approach that remains agnostic about the lens model \footnote{In \cite{Chakraborty:2024mbr, Chakraborty:2025maj, Chakraborty:2024net}, another lens-model-agnostic approach, which differs from ours, is employed. In this approach, the procedure involves obtaining the residuals of the gravitational wave (GW) data of multiple detectors by subtracting the best-fit unlensed waveform from the observed data. These residuals are then cross-correlated to determine the presence of a microlensing signal.}. Here, the amplification factor is calculated by dividing the observed waveform by the assumed unlensed waveform, which we do not know a priori. The advantage of this method, if successful, is that it does not require prior assumptions about the lens properties. However, a critical limitation arises: if the assumed unlensed waveform is incorrect, the resulting amplification factor is also incorrect and fails to represent any physically plausible lens model. Note that this issue is unrelated to measurement errors in gravitational waves caused by detector sensitivity; it persists even under perfect GW measurements. Addressing this fundamental issue by making use of the Kramers-Kronig (KK) relation is the central focus of this paper.

The KK relation is a well-known principle in optics \cite{lucarini2005kramers}, representing a general relationship satisfied by response functions in linear response systems governed by the principle of causality. As demonstrated in \cite{Tanaka:2023mvy}, the GL system falls into this category, and the KK relation, which links the real and imaginary parts of the amplification factor, holds for any lens model. If the amplification factor is incorrectly determined from observations, it generally does not satisfy the KK relation. Conversely, verifying whether the observed amplification factor is consistent with the KK relation provides a robust method to identify and dismiss false amplification factors without relying on specific lens models. This approach assumes that GWs are perfectly observed across the entire frequency range required for an exact evaluation of the KK relation. However, in practice, GW measurements are limited to a finite frequency range, causing the KK relation to appear violated even when the amplification factor is correctly determined. The purpose of this paper is to provide a proof-of-principle demonstration of the KK relation's effectiveness under the constraint of a limited observable frequency range.

The paper is organized as follows. In Sec. \ref{Sec: Review}, we briefly review the GL and KK relations and introduce the basic knowledge required in the subsequent sections. In particular, Sec. \ref{Sec: Review KK relation} and \ref{Sec: Violation truncation} address our previous work\cite{Tanaka:2023mvy}. In Sec. \ref{Sec: Methodology}, we describe our proposed methodology for applying the KK relation. First, we explain the violation of the KK relation, which is the heart of the methodology, in Sec. \ref{Sec: Violation false}, and then describe the specific analysis procedure in detail in Sec. \ref{Sec: Dismiss parameters}. In Sec. \ref{Sec: Results}, we present simulation results of the proposed methodology. We performed simulations for both cases where the observed waveforms are GL signals and where they are not. The results show that we can restrict the parameter space of the GW template by eliminating false amplification factors, and that under certain conditions it is possible to distinguish between GL and non-GL signals. Lastly, we use a unit with $c=1$.

\section{Review: Kramers-Kronig relation in gravitational lensing}
\label{Sec: Review}
In this section, we give an overview of the KK relation in the GL system following \cite{Tanaka:2023mvy}. 
The KK relation is a general relation for response function in a linear response system and is satisfied when the system has causality. 
As investigated in \cite{Tanaka:2023mvy}, the GL system can also be considered as a linear response system that obeys the causality. 
After introducing the amplification factor which is the response function in the
GL system, we confirm that the GL system satisfies several requirements and write down the KK relation in the GL system. 
Then, we introduce and quantify the violation of the KK relation caused by restricting  
to observable frequency range.

\subsection{Gravitational lensing}
\label{Sec: Review GL}
GL is a phenomenon in which gravity distorts the propagation of waves. 
In this paper, we focus on the propagation of GWs in the presence of a static and localized massive object. Furthermore, we ignore the change of polarization during propagation as sufficiently small and treat GWs as scalar waves\cite{schneider1999gravitational,peters1974index}. Then the propagation of GWs is described by the wave equation on the time-independent background Newtonian potential\footnote{Here we take into account the cosmic expansion. However, since the wavelength of GWs is sufficiently shorter than the horizon scale, the wave equation Eq. (\ref{wave eq}) is the same as for flat spacetime using the comoving coordinate. All we have to do is replace $\omega$ and the distances in the flat spacetime amplification factor with $(1+z_L)\omega$ and angular diameter distances.}:
\begin{eqnarray}
    (\nabla^2 + \omega^2)\phi_L(\omega, \bm{r})=4\omega^2\Phi(\bm{r})\phi_{L}(\omega, \bm{r})~,\label{wave eq}
\end{eqnarray}
where $\phi_{L}$ represents lensed GWs. To represent waveform distortion due to the GL effect, we introduce the amplification factor $F(\omega)$ which is defined by
\begin{eqnarray}
    \phi_L(\omega, \bm{r}_{\rm{obs}}) = F(\omega) \phi_0(\omega, \bm{r}_{\rm{obs}})~,\label{def of ampfac}
\end{eqnarray}
where $\bm{r}_{\rm{obs}}$ is the position of the observer and $\phi_0$ is unlensed GWs which is the solution of Eq. (\ref{wave eq}) when $\Phi=0$. In terms of $F(\omega)$, we can solve Eq. (\ref{wave eq}), and under the thin-lens approximation\cite{Suyama:2005mx}\footnote{For simplicity, the thin-lens approximation is used in this paper. However, the KK relation holds without the approximation, and our methodology described in Sec. \ref{Sec: Methodology} is valid.}, the expression for $F(\omega)$ is obtained as
\begin{equation}
    F(w)=\frac{w}{2\pi i}\int d^2x~e^{iw[T(\bm{x},\bm{y})-T_{\rm{min}}(\bm{y})]}~,\label{diffraction integral}
\end{equation}
\begin{equation}
    T(\bm{x}, \bm{y})=\frac{1}{2}(\bm{x}-\bm{y})^2-\psi(\bm{x}),~~~T_{\rm{min}}(\bm{y})=\min_{\bm{x}}T(\bm{x},\bm{y})~.
\end{equation}
\begin{figure}
    \centering
    \includegraphics[scale=0.4]{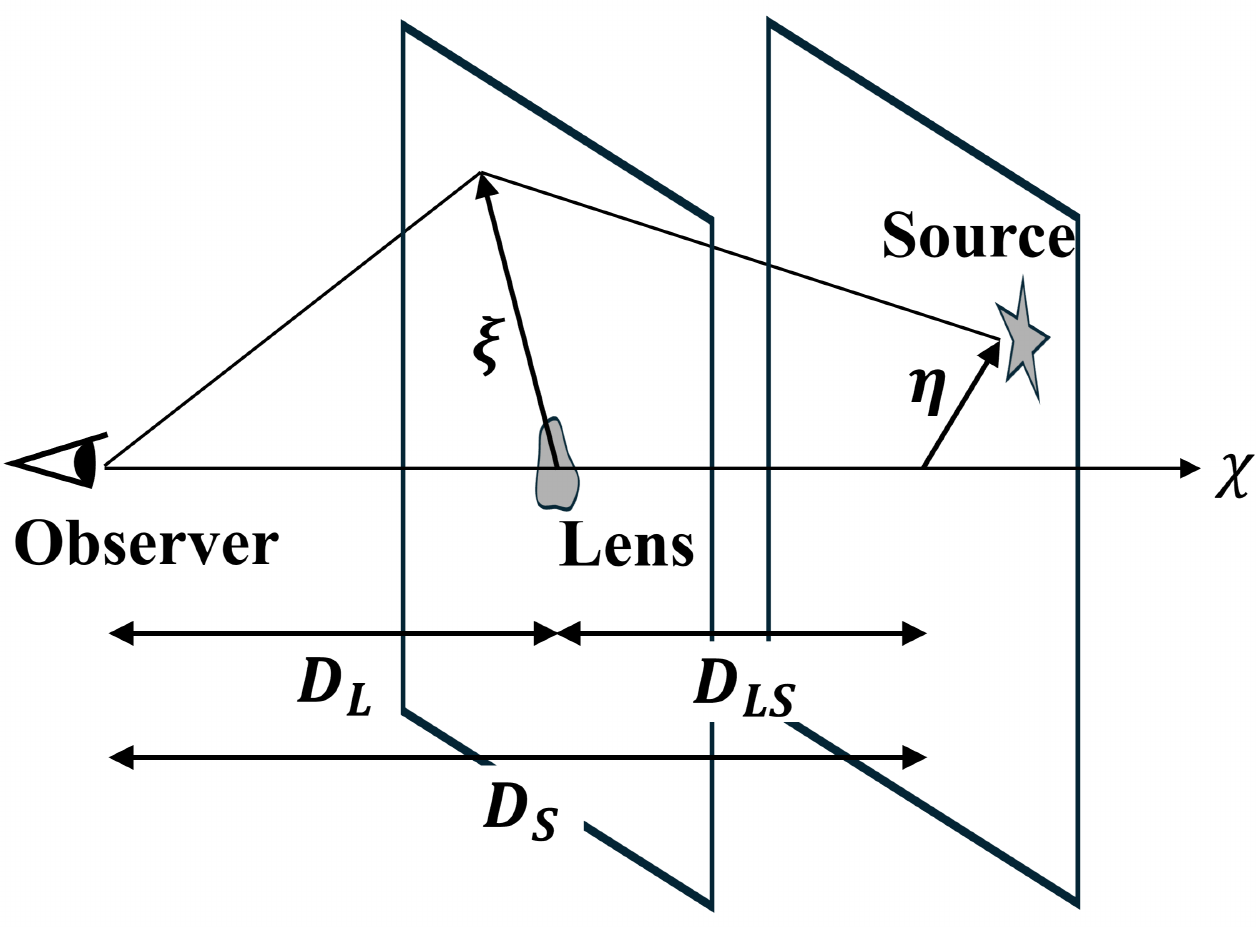}
    \caption{Schematic picture of GL. $\bm{\xi}$ and $\bm{\eta}$ are the position vectors in the lens and the source plane, respectively. $D_L$ is the angular diameter distance between the observer and the lens, $D_{LS}$ is between the lens and the source, and $D_S$ is between the observer and the source.}
    \label{Fig: GL setup}
\end{figure}
Here we introduced dimensionless quantities $\bm{x}, \bm{y}, w$ by
\begin{eqnarray}
    \bm{x}=\frac{\bm{\xi}}{D_L\theta_*},~~\bm{y}=\frac{\bm{\eta}}{D_S\theta_*},~~w=\theta_*^2\frac{D_LD_S}{D_{LS}}(1+z_L)\omega~,
\end{eqnarray}
where $\bm{\xi}$ is the position in the lens plane, $\bm{\eta}$ is the position of the source in the source plane, $D_L$ is the angular diameter distance between the observer and the lens, $D_{LS}$ is between the lens and the source, and $D_S$ is between the observer and the source (see Fig. \ref{Fig: GL setup}). $z_L$ is the redshift of lens and $\theta_*$ is an normalization angular constant to simplify the expression of the lens potential.
Moreover, $\psi(\bm{x})$ is the dimensionless lens potential defined by
\begin{eqnarray}
    \psi(\bm{x}) = \theta_*^{-2}\frac{D_{LS}}{D_LD_S}\int_{0}^{D_s} d\chi~2\Phi(\chi,\bm{\xi})
\end{eqnarray}
which satisfies the two-dimensional poison equation
\begin{eqnarray}
    \nabla^2\psi(\bm{x})=2\kappa(\bm{x})~,
\end{eqnarray}
where $\chi$ is the comoving distance from the observer, and $\kappa(\bm{x})$ is the dimensionless version of the surface mass density distribution $\Sigma(\bm{\xi})$:
\begin{eqnarray}
    \kappa(\bm{x})=\frac{\Sigma(\bm{\xi})}{\Sigma_{\rm{cr}}},~~\Sigma_{\rm{cr}}=\frac{D_S}{4\pi GD_LD_{LS}}~.
\end{eqnarray}
In the following, we consider $\Sigma(\bm{\xi})$ in two models and obtain the amplification factor from Eq. (\ref{diffraction integral}).

\subsubsection{Point mass lens}
\begin{figure}
    \centering
    \begin{minipage}{0.47\columnwidth}
        \centering
        \includegraphics[width=1\columnwidth]{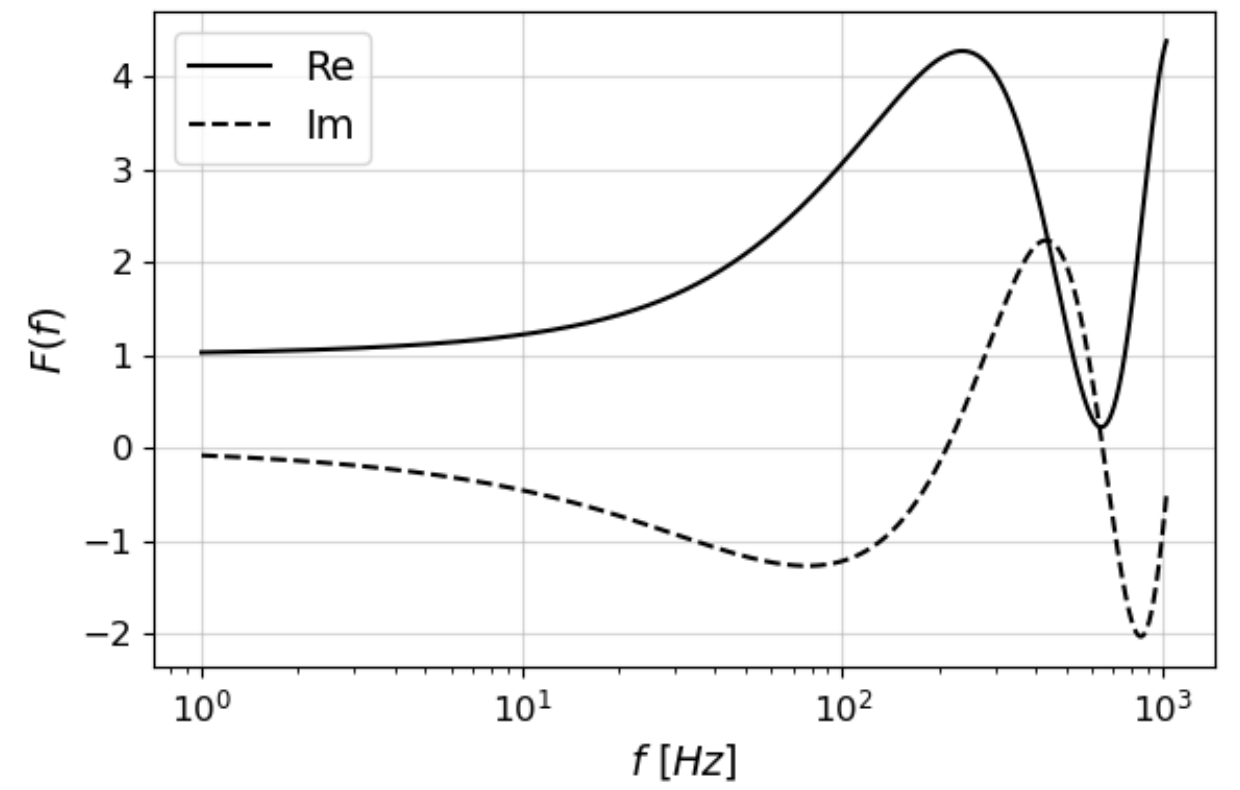}
    \end{minipage}
    \begin{minipage}{0.47\columnwidth}
        \centering
        \includegraphics[width=1\columnwidth]{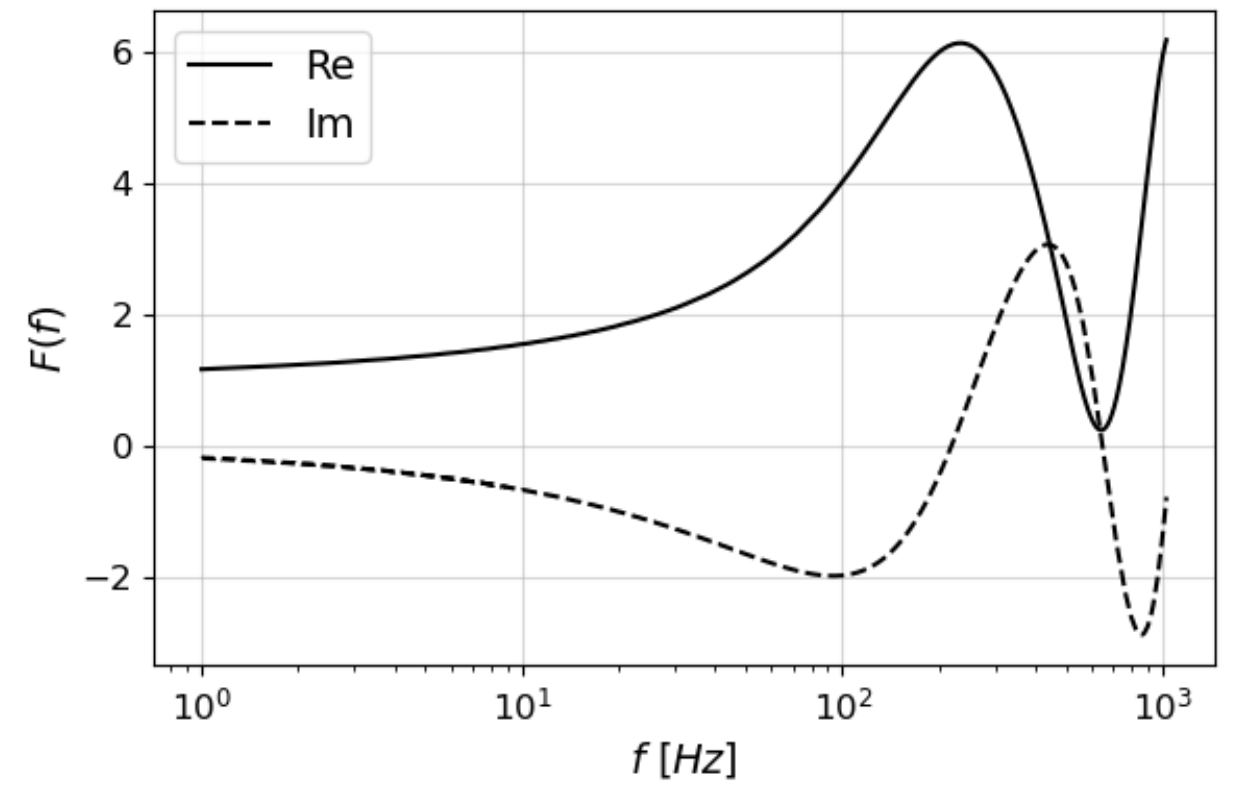}
    \end{minipage}
    \caption{The amplification factors in the PML (left panel) and SIS models (right panel). 
    The solid lines are the real parts and the dotted lines are the imaginary parts. In both cases, the lens parameters are $M_{Lz}=300M_{\odot},~y=0.1$.}
    \label{Fig: ampfac_PML_SIS}.
\end{figure}

The simplest lens model is the Point Mass Lens (PML) model. This model is characterized by the surface mass density
\begin{eqnarray}
    \Sigma(\bm{\xi})=M_L\delta^2(\bm{\xi})~,
\end{eqnarray}
where $M_L$ is the lens mass. 
Here we normalize by $\theta_*=(4GM_LD_{LS}/D_LD_S)^{1/2}$. In this normalization, $w=4GM_{Lz}\omega$ where $M_{Lz}=(1+z_{L})M_{L}$ is the redshifted lens mass. The dimensionless surface mass density is 
\begin{eqnarray}
    \kappa(\bm{x})=\pi \delta^2(\bm{x})~,
\end{eqnarray}
and thus the lens potential is
\begin{eqnarray}
    \psi(\bm{x})=\ln x~.
\end{eqnarray}
In this case, Eq. (\ref{diffraction integral}) can be analytically integrated\cite{Nakamura:1999uwi}:
\begin{eqnarray}
    F(w)=\exp\qty[\frac{\pi w}{4}+iw\qty(\frac{1}{2}\ln\qty(\frac{w}{2})-T_{\rm{min}})]\Gamma\qty(1-\frac{iw}{2}){}_1F_1\qty(\frac{iw}{2},1;\frac{iwy^2}{2})~,
\end{eqnarray}
where 
\begin{equation}
    T_{\rm{min}}=\frac{2}{(y+\sqrt{y^2+4})^2}-\ln\qty(\frac{y+\sqrt{y^2+4}}{2})~.
\end{equation}
The amplification factor for $M_{Lz}=300M_{\odot},~y=0.1$ is shown in the left panel of Fig. \ref{Fig: ampfac_PML_SIS}.

\subsubsection{Singular isothermal sphere}
Singular Isothermal Sphere (SIS) is a model that represents the mass distribution of galaxies\cite{schneider1999gravitational}, 
which is given by
\begin{eqnarray}
    \Sigma(\bm{\xi})=\frac{\sigma_v^2}{2G\xi}~,
\end{eqnarray}
where $\sigma_v$ is the velocity dispersion of stars in the galaxy. 
Here we choose the normalization constant to be $\theta_* = 4\pi\sigma_v^2D_{LS}/D_S$. 
Then the dimensionless surface mass density is
\begin{eqnarray}
    \kappa(\bm{x})=\frac{1}{2x}
\end{eqnarray}
and the lens potential is
\begin{eqnarray}
    \psi(\bm{x})=x~.
\end{eqnarray}
Here we define the lens mass as the mass inside the Einstein radius $\xi_E=4\pi\sigma_v^2D_LD_{LS}/D_{S}$, and thus $GM_L=4\pi^2\sigma_v^4D_LD_{LS}/D_S$. By defining the mass in this way, the expression for the normalized frequency becomes the same as for the PML, i.e., $w=4GM_{Lz}\omega$. In this model, Eq. (\ref{diffraction integral}) can not be analytically integrated, therefore we integrate it numerically\cite{Tambalo:2022plm}. The amplification factor for $M_{Lz}=300M_{\odot},~y=0.1$ is shown in the right panel of Fig. \ref{Fig: ampfac_PML_SIS}.

\subsection{Kramers-Kronig relation}
\label{Sec: Review KK relation}
The KK relation is satisfied when a linear response system has causality. From Eq. (\ref{def of ampfac}), the GL system can be regarded as a linear response system with $\phi_0$ as input, $\phi_L$ as output, and $F$ as response function. Furthermore, the GL system has the causality that any waves will reach the observer at the same time or later than the light if they are emitted simultaneously from the wave source.  
To be more precise, waves in the short wavelength limit choose the path that arrives earliest, as can be seen from Fermat's principle, so waves with finite wavelengths arrive at the same time or later\cite{Suyama:2020lbf,Tanaka:2023mvy}.
Based on this observation, in \cite{Tanaka:2023mvy} the KK relation for the amplification factor 
was derived. Its form is given by
\begin{equation}
    F(\omega)-1=\frac{\omega}{\pi i}\dashint_{-\infty}^{\infty}\frac{d\omega'}{\omega'-\omega}\frac{F(\omega')-1}{\omega'}~,\label{KK in GL}
\end{equation}
or by using $K\equiv\Re F-1,~S\equiv\Im F$ and taking the real part of the above, we get
\begin{equation}
    K(\omega)=\frac{\omega}{\pi}\dashint_{-\infty}^{\infty}\frac{d\omega'}{\omega'-\omega}\frac{S(\omega')}{\omega'}~,\label{KK in GL KS}
\end{equation}
where the integral symbol appearing in the above equations represents Cauchy's principal value integral. 
As evident from Eqs. (\ref{KK in GL}) and (\ref{KK in GL KS}), 
the KK relation links the real part of the amplification factor at a specific frequency to the imaginary part across all frequencies, and vice versa. 
Notably, Eq. (\ref{KK in GL}) is universally valid for any lens.
This naturally motivates us to make use of KK relation to test the observationally claimed lensed signal
without relying on specific lens models.

\subsection{Violation of the KK relation due to finite frequency range}
\label{Sec: Violation truncation}

In an ideal situation where the GW observation has been done for all frequencies (without measurement errors), 
KK relation is violated only when the measured amplification factor does not represent the correct lens signal.
In reality, the frequency range is limited and the KK relation is violated even when the measured amplification
factor is true, which must be accounted for to evaluate the effectiveness of the KK relation.
In our methodology, we focus on the violation of the KK relation caused by truncating the frequency range
to finite width. Therefore, in this subsection, 
we briefly review the violation of the KK relation caused by the GW observations limited to a finite frequency range. In real-world GW observations, the observable frequency range is limited. It is determined by the noise power spectrum of detectors, e.g., $10\--10^3~\text{[Hz]}$ for LIGO\cite{LIGOScientific:2014pky} and $10^{-3}\--10^{-1}~\text{[Hz]}$ for LISA\cite{Babak:2021mhe}. In reality, however, it is not possible to define the frequency range so clearly from the noise power spectrum. 
Because our purpose is a proof-of-principle demonstration, here we consider an ideal situation, where the noise is zero in some frequency range and infinite in others.
Then taking the observable frequency range as $O\equiv\{f\mid f_{\rm{min}}\leq |f| \leq f_{\rm{max}}\}$,  the range of frequencies outside of $O$ pertaining to the limits in the definite integral of Eq.(\ref{KK in GL KS}) must be truncated as:
\begin{equation}
    K(f)\neq\frac{f}{\pi}\dashint_O\frac{df'}{f'-f}\frac{S(f')}{f'}~,\label{KK in GL KS truncated}
\end{equation}
where $f=\omega/2\pi$ is used instead of $\omega$ for later convenience. Of course, since the equation has been modified, the KK relation is not satisfied, and we define $\Delta_{\rm{tr}}$ as quantifying the degree of the violation of the KK relation:
\begin{eqnarray}
    \Delta_{\rm{tr}}(f)&\equiv& K(f)-\frac{f}{\pi}\dashint_{O}\frac{df'}{f'-f}\frac{S(f')}{f'}\nonumber\\
    &=&\frac{f}{\pi}\dashint_{\overline{O}}\frac{df'}{f'-f}\frac{S(f')}{f'}~,\label{Delta tr}
\end{eqnarray}
where $\overline{O}=[-\infty,\infty]-O$ and we used Eq. (\ref{KK in GL KS}) from the first to the second line. 

\begin{figure}[t]
    \centering
    \includegraphics[scale=0.45]{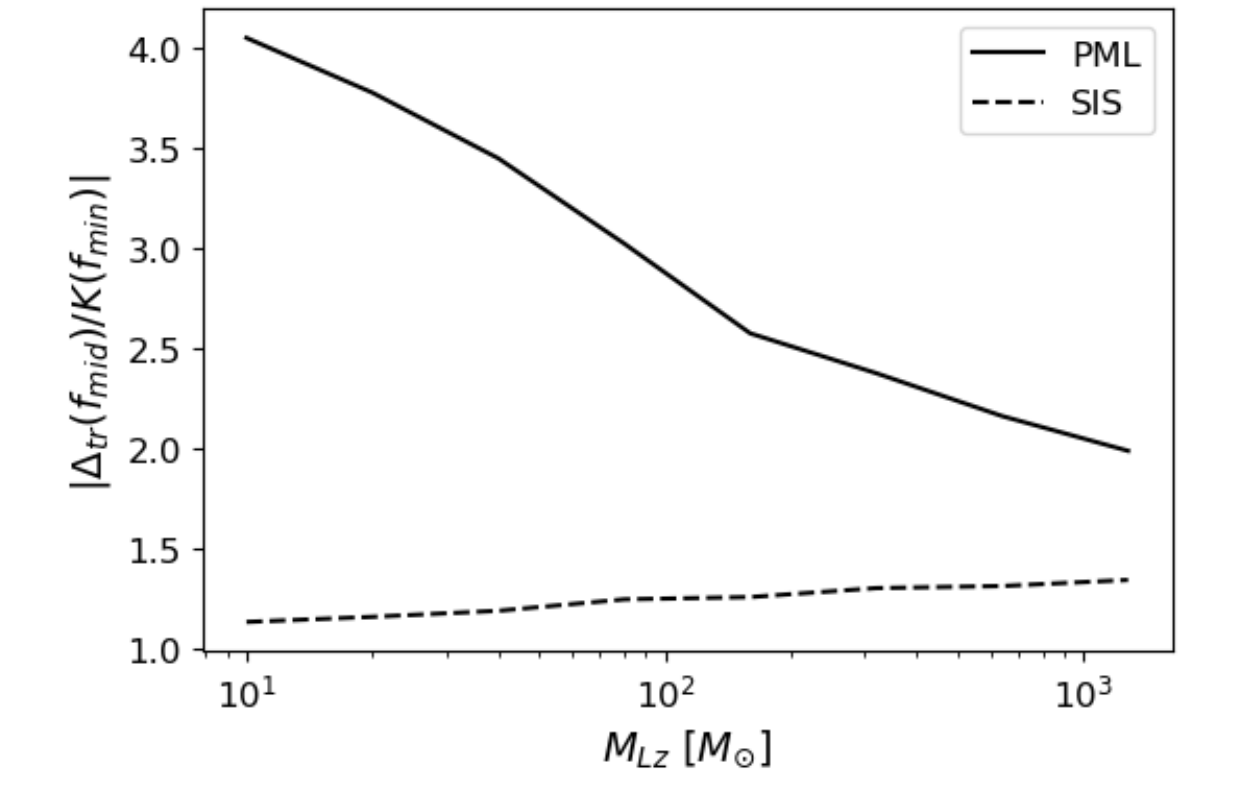}
    \caption{Plot of the $|\Delta_{\rm{tr}}(f_{\rm{mid}})/K(f_{\rm{min}})|$ in Eq. (\ref{estimation}). The solid line is for PML and the dotted line is for SIS. The parameters are $y=0.1,~f_{\rm{min}}=1\text{ [Hz]},~f_{\rm{max}}=1024\text{ [Hz]}$. In this parameter region, we can confirm that the $|\Delta_{\rm{tr}}(f_{\rm{mid}})/K(f_{\rm{min}})|$ is indeed $\mathcal{O}(1)$ as expected in Eq. (\ref{estimation}). The reason we stopped $M_{Lz}$ at $10M_{\odot}$ is that below this value, the amplification factor is in the WO regime in $[f_{\rm{min}}, f_{\rm{max}}]$, so the condition (\ref{hierarchy}) is not satisfied and is not our target as explained in the last paragraph of Sec. \ref{Sec: Dismiss parameters}. 
    See Appendix for an discussion of why $|\Delta_{\rm{tr}}(f_{\rm{mid}})/K(f_{\rm{min}})|$ increases as $M_{Lz}$ decreases.}
    \label{Fig: O(1) coefficient}
\end{figure}

As discussed in \cite{Tanaka:2023mvy}, in the middle of the frequency range $f_{\rm{min}}\ll f \ll f_{\rm{max}}$, $\Delta_{\rm{tr}}(f)$ is almost constant. 
Furthermore, when $f_{\rm{min}}$ is sufficiently low and in the WO regime\footnote{
We define the WO regime as the range where the amplification factor deviates only slightly from unity, allowing the Born approximation to provide a reasonably accurate estimate of the amplification factor. 
For instance, for the point-mass lens, $f$ is in the WO regime when $f\ll1/GM_{Lz}\simeq10^5 M_{\odot}/M_{Lz}\text{~[Hz]}$.} that the Born approximation\cite{Mizuno:2022xxp} can be used, we can estimate the value of the constant as
\begin{eqnarray}
    |\Delta_{\rm{tr}}(f_{\rm{mid}})| = \mathcal{O}(1)\times |K(f_{\rm{min}})|~,\label{estimation}
\end{eqnarray}
where $f_{\rm{min}}\ll f_{\rm{mid}}=\sqrt{f_{\rm{min}}f_{\rm{max}}}\ll f_{\rm{max}}$\footnote{The reason we define $f_{\rm mid}$ as a multiplicative average of $f_{\rm min}$ and $f_{\rm max}$ rather than an additive average is to ensure that $f_{\rm{min}}\ll f_{\rm{mid}}\ll f_{\rm{max}}$ is satisfied when $f_{\rm{min}}\ll f_{\rm{max}}$}. and we assume the frequency range is broad enough so that there is a large separation between $f_{\rm min}$ and $f_{\rm mid}$.
It should be emphasized here that Eq.~(\ref{estimation}) is valid for any lens models and the information of the lens is encoded in the ${\cal O}(1)$ coefficient.
As shown in Fig.~\ref{Fig: O(1) coefficient}, we explicitly confirm in some parameter region that this estimation is in fact correct in the PML and SIS models. 
In particular, we find that $\Delta_{\rm{tr}}(f_{\rm{mid}})/|K(f_{\rm{min}})|$ for SIS models is smaller
than that for PML. 
As we argue in Appendix, $\Delta_{\rm{tr}}(f_{\rm{mid}})/|K(f_{\rm{min}})|$ for other lens models is also 
smaller than that for PML.
Thus, the relation (\ref{estimation}) holds true when the amplification factor is true and is in the WO regime
at $f_{\rm{min}}$.
In this context, (\ref{estimation}) can be seen as the finite-frequency analogue of the KK relation
and plays a central role in our methodology explained in the next section.

\section{Methodology to dismiss false amplification factor}
\label{Sec: Methodology}
In GL observations using GWs, the observable quantity is the GW waveform $\phi_{\rm{obs}}(f)$. $\phi_{\rm{obs}}(f)$ may or may not be a waveform affected by the GL effect. If we assume that $\phi_{\rm{obs}}(f)$ is a lensed waveform, the key task is to correctly extract the amplification factor from it or determine whether no GL signal is present.
As we have mentioned in the Introduction, we take an agnostic approach to the lens properties (i.e., no templates for the lens).
To obtain the amplification factor, we need to select a template waveform $\phi_{\rm{temp}}(f)$ as an unlensed waveform (which cannot be directly observed), and the amplification factor $F_{\rm{obs}}(f)$ is obtained as the modulation of the observed waveform from the unlensed waveform based on the definition in Eq. (\ref{def of ampfac}):
\begin{equation}
    F_{\rm{obs}}(f)=\frac{\phi_{\rm{obs}}(f)}{\phi_{\rm{temp}}(f)}~.\label{Fobs}
\end{equation}
Thus, when we simply refer to the template below, it refers to the template used for the unlensed waveform.

However, the process described above Eq. (\ref{Fobs}) may yield false amplification factors\footnote{
When we refer to a true amplification factor, we mean an amplification factor that correctly represents the GL effect that is actually occurring. Conversely, one that does not is called a false amplification factor.
}, which can be categorized into the following two cases.
The first case is when the observed waveform is really a GL signal, i.e., a lensed waveform. 
In this case, if the model or parameters of the chosen template waveform are false, we obtain the false amplification factor.
The second case is when the observed waveform is not a GL signal, i.e., an unlensed waveform not covered by templates and is misidentified as a GL signal. 
An example is when the observed waveform is binary GWs from elliptical orbit and template for the source is assumed to be GWs from quasi-circular orbit.
In this case, there is a possibility that the GL analysis could misidentify GWs emitted from elliptical orbit as a GL signal \footnote{A detailed study probing the degeneracy between microlensed and eccentric GW signals, and the resulting systematics, is planned to be reported soon [A. Mishra et al (in prep)].}. In addition, in principle, GWs from sources that have not yet been modeled could be misidentified as GL signals, leading to false amplification factors. As in the two cases above, the observation may yield false amplification factors. 
In this section, we will explain our methodology to dismiss such false amplification factors using the KK relation.

We first define the violation of the KK relation due to a false amplification factor in Sec. \ref{Sec: Violation false}.
In that subsection, we provide a grounded argument for dismissing false amplification factors, using the estimation of $\Delta_{\rm{tr}}$ in Sec. \ref{Sec: Violation truncation}. In Sec. \ref{Sec: Dismiss parameters}, we explain the detailed procedure of our methodology.

\subsection{Violation of the KK relation due to false amplification factor}
\label{Sec: Violation false}

\begin{figure}[t]
    \centering
    \begin{minipage}{0.47\columnwidth}
        \centering
        \includegraphics[width=1\columnwidth]{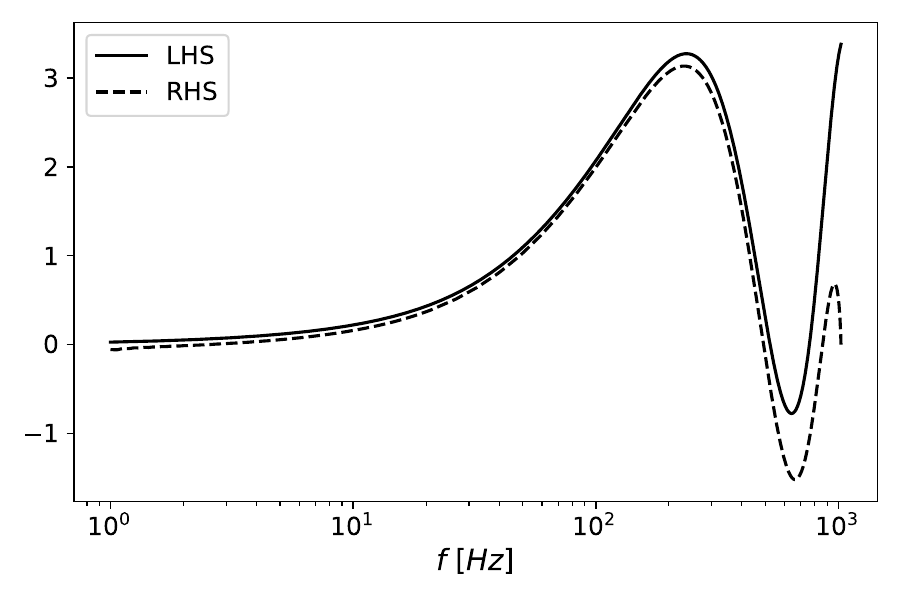}
    \end{minipage}
    \begin{minipage}{0.47\columnwidth}
        \centering
        \includegraphics[width=1\columnwidth]{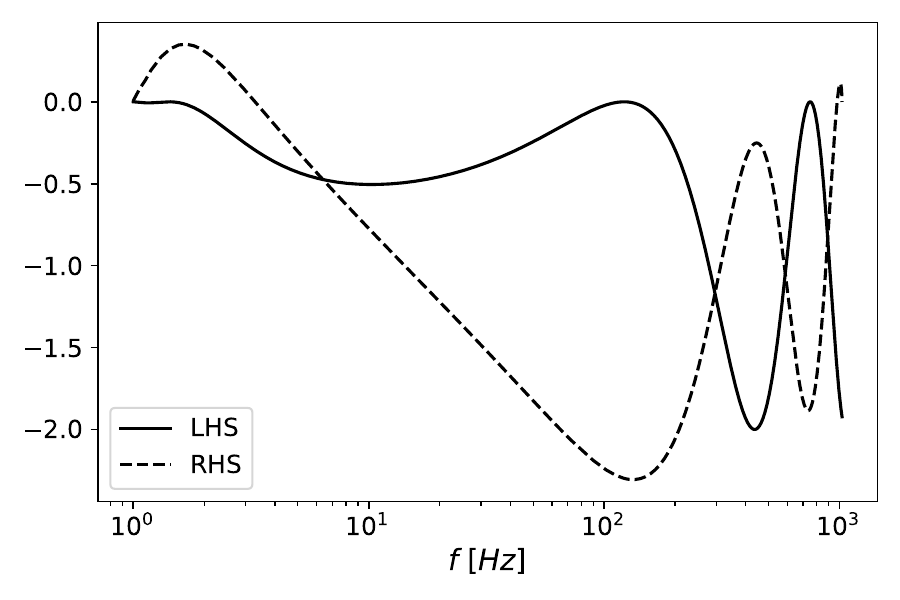}
    \end{minipage}
    \caption{Comparison of the left-hand side (solid line) and right-hand side (dashed line) of Eq. (\ref{KK in GL KS truncated}) with frequency range $[1~\text{Hz},~1024~\text{Hz}]$. The left panel is for the true amplification factor in the PML model with $M_{Lz}=300M_{\odot},~y=0.1$. In this case, the KK relation is slightly violated only due to the limited frequency range. The right panel is for the false amplification factor obtained from GWs with the eccentric orbit (see Sec. {\ref{Sec: ECC signal}}). The parameters are $e = 0.01,~\bm{\theta}_0=\{30M_{\odot},~1000~\text{Mpc},~0~\text{s},~0~\text{rad}\},~\bm{\theta}=\{30.005M_{\odot},~1000~\text{Mpc},~0.01~\text{s},~-0.01~\text{rad}\}$. For this false amplification factor, the KK relation is obviously violated, indicating that the violation is not only due to the limited frequency range.} 
    \label{Fig: LHS_RHS}
\end{figure}

 A false amplification factor generally violates the KK relation as shown in Fig. \ref{Fig: LHS_RHS}. In this subsection, we discuss the criterion for the magnitude of the violation required to dismiss false amplification factors considering the limited frequency range.
Let $F_{\rm{obs}}$ denote the observationally inferred amplification factor which is not necessarily a true amplification factor and define the quantity that quantifies the violation of the KK relation as
\begin{equation}
    \Delta_{\rm{obs}}(f)\equiv K_{\rm{obs}}(f)-\frac{f}{\pi}\dashint_{O}\frac{df'}{f'-f}\frac{S_{\rm{obs}}(f')}{f'}~.\label{Delta}
\end{equation}
Here, the integral range is truncated because of the limited frequency range in observations as described in Sec. \ref{Sec: Violation truncation}. When $F_{\rm{obs}}$ is the true amplification factor, i.e., it correctly represents physical phenomenon, then there is only a contribution to $\Delta_{\rm{obs}}(f)$ from the truncation, and $\Delta_{\rm{obs}}(f)=\Delta_{\rm{tr}}(f)$.
Furthermore, when $F_{\rm{obs}}$ is the true amplification factor with $f_{\rm{min}}$ in the WO regime, 
according to the discussion in Sec. \ref{Sec: Violation truncation},
we have 
\begin{equation}
    |\Delta_{\rm{obs}}(f_{\rm{mid}})|= \mathcal{O}(1)\times |K_{\rm{obs}}(f_{\rm{min}})|~.\label{estimation obs}
\end{equation}
Conversely, if $\Delta_{\rm{obs}}(f_{\rm{mid}})$ does not satisfy this relation, we can assert that $F_{\rm{obs}}$ is not the true amplification factor with the WO regime at the low-frequency side. 
Based on this consideration, by introducing a quantity $r$ as 
\begin{equation}
    r\equiv\qty|\frac{\Delta_{\rm{obs}}(f_{\rm{mid}})}{K_{\rm{obs}}(f_{\rm{min}})}|~, \label{r}
\end{equation}
our criterion for dismissing false amplification factors is stated as follows:
\begin{equation}
    \text{If}~r>r_{\rm{th}},~F_{\rm{obs}}~\text{is not the true amplification factor with the WO regime around $f_{\rm min}$}.\label{statement: r>O(1)}
\end{equation}
Here $r_{\rm{th}}$ is an $O(1)$ threshold value set by hand.
The above statement is the basis of our methodology and the rationale for dismissing false amplification factors, as we use later.

\subsection{Procedure to dismiss false amplification factor}
\label{Sec: Dismiss parameters}

In this subsection, we describe the detailed procedures of our methodology.
Let $\phi_{\rm{obs}}(f)$ be an observed waveform in the frequency range
$[f_{\rm min}, f_{\rm max}]$ with $f_{\rm max} \gg f_{\rm min}$. First, we make the following 
hypothesis: 
\begin{eqnarray}
    &&\text{$\phi_{\rm{obs}}$ is a lensed waveform in the WO regime around $f_{\rm min}$}\nonumber\\
    &&\text{with an unlensed waveform $\phi_{\rm{temp}}$}.
    \label{assumption: with WO}
\end{eqnarray}
Here $\phi_{\rm{temp}}$ is a template waveform calculated from some source model.
That the low-frequency side is in the WO regime is necessary for the use of the estimation of $\Delta_{\rm{obs}}$ discussed in Sec. \ref{Sec: Violation truncation} and Sec. \ref{Sec: Violation false}. 
Next, we obtain the would-be amplification factor as
\begin{equation}
    F_{\rm{obs}}(f;\bm{\theta})=\frac{\phi_{\rm{obs}}(f)}{\phi_{\rm{temp}}(f;\bm{\theta})}~,
\end{equation}
where $\bm{\theta}$ is a set of parameters of the template waveform. 
Then, to see if $F_{\rm{obs}}$ violates the relation (\ref{estimation obs}), we compute
\begin{equation}
    r(\bm{\theta})=\qty|\frac{\Delta_{\rm{obs}}(f_{\rm{mid}};\bm{\theta})}{K_{\rm{obs}}(f_{\rm{min}};\bm{\theta})}|\label{r theta}
\end{equation}
as discussed in Sec. \ref{Sec: Violation false}. 
If $r(\bm{\theta})$ is greater than $r_{\rm th}$, we decide from the statement $(\ref{statement: r>O(1)})$ that $F_{\rm{obs}}(f;\bm{\theta})$ is false and $\bm{\theta}$ is not the true set of parameters. 
In this way, our methodology can dismiss false parameters and restrict the template's parameter space. Furthermore, if $r(\bm{\theta})>r_{\rm{th}}$ for any $\bm{\theta}$, i.e., 
if no parameters satisfy the relation (\ref{estimation obs}), we conclude that Hypothesis (\ref{assumption: with WO}) is false. In that case, $\phi_{obs}$ is either not a GL signal or is a GL signal but without the WO regime or is a GL signal with an unlensed template not used in the analysis. In this way, our methodology allows us to test Hypothesis (\ref{assumption: with WO}) and to dismiss it if it is false.

\begin{figure}[t]
    \centering
    \includegraphics[scale=0.5]{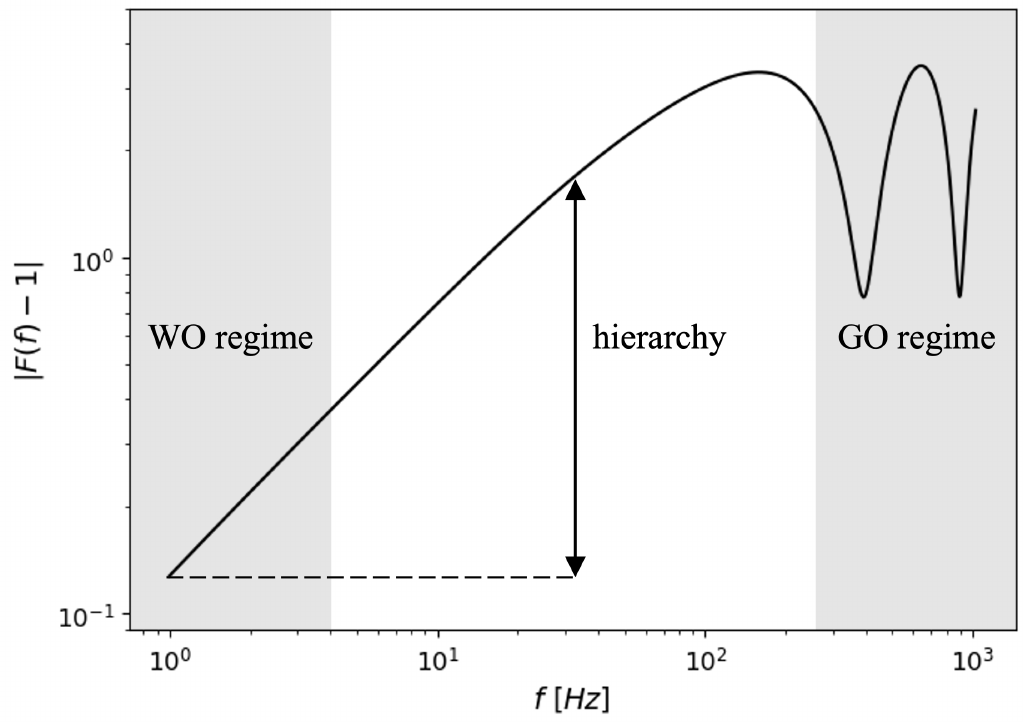}
    \caption{Schematic illustration of the hierarchy. This is the amplification factor of the SIS model with $M_{Lz}=500M_{\odot},~ y=0.1$. The shaded areas on the left and right are the WO and GO regimes, respectively. For the KK relation to be valid, there must be a sufficient hierarchy between the values of the amplification factor at $f_{\rm{mid}}$ and $f_{\rm{min}}$, as shown in the figure.}
    \label{Fig: hierarchy}
\end{figure}

We perform the above procedure while varying $\bm{\theta}$ but restricting the procedure only to the region of $\bm{\theta}$ where the obtained amplification factor
at $f_{\rm min}$ is in the WO regime, as required by Hypothesis (\ref{assumption: with WO}).
To quantify this restriction, we introduce a quantity, which we call hierarchy, defined by 
\begin{equation}
    h(\bm{\theta})\equiv\qty|\frac{F_{\rm{obs}}(f_{\rm mid};\bm{\theta})-1}{F_{\rm{obs}}(f_{\rm min};\bm{\theta})-1}|\label{hierarchy def}
\end{equation}
and we confine our analysis to the parameter space of $\bm{\theta}$ to $D$ which is defined by
\begin{equation}
    D\equiv\{\bm{\theta}\mid h(\bm{\theta})>h_{\rm{th}}\}~,\label{hierarchy}
\end{equation}
where $h_{\rm{th}}$ is a threshold of the hierarchy set by hand to exclude from our analysis any false amplification factors that 
do not resemble the WO signal at $f_{\rm min}$. Specifically, amplification factors such as the one shown in Fig. \ref{Fig: hierarchy}, where $f_{\rm min}$ is in the WO regime and $f_{\rm max}$ is in the GO regime and satisfy the condition of Eq. (\ref{hierarchy}) are included in our analysis
\footnote{
Since the condition for $f$ being in the WO regime is $f\ll10^5 M_{\odot}/M_{Lz}\text{~[Hz]}$, the lens mass of the amplification factor included in our analysis is $10^5M_{\odot}\times\text{Hz}/f_{\rm max}\lesssim M_{Lz}\ll10^5M_{\odot}\times \text{Hz}/f_{\rm min}$.
\label{footnote: mass range}}
, while amplification factors where both $f_{\rm min}$ and $f_{\rm max}$ are in the WO regime and do not satisfy the condition of Eq. (\ref{hierarchy}) are not included.

To determine the appropriate value for $h_{\rm th}$, let us consider a false amplification factor $F_{\rm obs}(f;\bm{\theta})$ that satisfies $r(\bm{\theta})\le r_{\rm th}$ but 
does not resemble the WO signal
at $f_{\rm min}$.
For the true amplification factor, cancellation between the first and the second term on the right-hand side of Eq. (\ref{Delta})
occurs as a consequence of the KK relation and
the right-hand side of Eq. (\ref{Delta}) becomes as small as $K(f_{\rm min}) (\ll K(f_{\rm mid}))$ (see Fig.~\ref{Fig: hierarchy}). 
On the other hand, such cancellation does not happen for the false amplification factors in general. 
As a result, $|\Delta_{\rm obs}(f_{\rm mid})|$ is just comparable to the amplification factor at the same
frequency: $|\Delta_{\rm obs}(f_{\rm mid})|\simeq|F_{\rm obs}(f_{\rm mid})-1|$.
Then, by the definition of the hierarchy $h$, we have
\begin{equation}
|\Delta_{\rm obs}(f_{\rm mid})|\simeq \bigg| \frac{F_{\rm{obs}}(f_{\rm mid})-1}{F_{\rm{obs}}(f_{\rm min})-1} \bigg|
|F_{\rm obs}(f_{\rm min})-1| 
\ge h |K(f_{\rm min})|,
\end{equation}
where we have used a general inequality $|F_{\rm obs}(f_{\rm min})-1| \ge |K(f_{\rm min})|$.
Comparing this relation with the definition of $r$ (\ref{r}), we find $h \le r (\le r_{\rm th})$.
Thus, by choosing $h_{\rm th}$ to be larger than $r_{\rm th}$, the false amplification factor under consideration always satisfies $h < h_{\rm th}$
and becomes outside the domain $D$.
Based on this observation, we set $h_{\rm th}=r_{\rm th}$ by which false amplification factors satisfying $r(\bm{\theta})\le r_{\rm th}$ with non-WO signal
at $f_{\rm min}$ are excluded from our analysis \footnote{It is preferable to keep $h_{\rm th}$ as small as possible, since a larger $h_{\rm th}$ narrows the range of GL signals to be targeted.}.

\begin{figure}[t]
    \centering
    \includegraphics[scale=0.35]{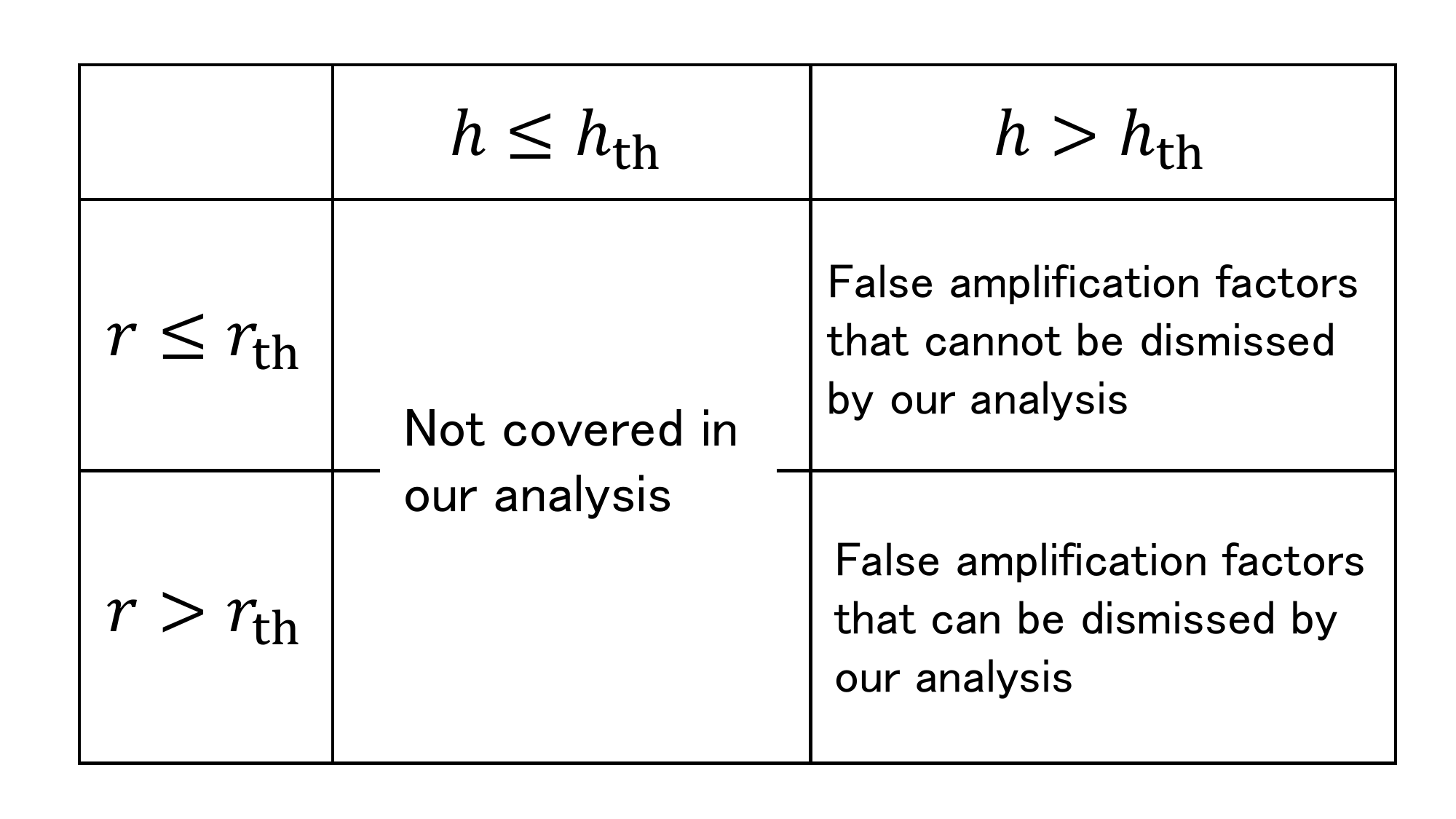}
    \caption{Table categorizing false amplification factors by the values of $r$ and $h$. The domain $h\leq h_{\rm th}$ is not covered by our analysis due to the condition on the hierarchy (\ref{hierarchy}). For the remaining domain $h > h_{\rm th}$,  the region with $r> r_{\rm th}$ 
    is where our method is effective and can dismiss any $F$ as false amplification factors by Criterion (\ref{statement: r>O(1)}). 
    On the other hand, the region with $r\leq r_{\rm th}$ can not be dismissed by our methodology; any false amplification factor belonging to this region mimics the
    true amplification factor from the point of view of the KK relation.}
    \label{Fig: Table-r-h}
\end{figure}

\section{Results}
\label{Sec: Results}

In this section, we perform simulations according to the methodology described in Sec. \ref{Sec: Methodology}. In Sec. \ref{Sec: True signal}, we check if our methodology gives consistent results when an observed waveform is a GL signal. In this case, a true set of parameters exists in the parameter space of the template waveform, and we confirm in the PML and SIS models that the set of those parameters is not dismissed by the KK relation. In Sec. \ref{Sec: False signal}, we investigate the effectiveness of our methodology when an observed waveform is a non-GL signal, and we check how many parameters can be dismissed by the KK relation and show that in some cases, the non-GL signal can be robustly dismissed.

In reality, GW template waveforms include many models of the GW source, such as GWs with eccentric binary orbit and GWs with the self-rotational spins of the individual components of the binary. In this section, however, we treat only non-spinning GWs from quasi-circular orbit as a template waveform for simplicity. Furthermore, the binary stars are assumed to be equal mass, and four parameters are taken as its parameters: redshifted mass of a star $m$, luminosity distance from the observer $d$, coalescence time $t_{\rm{coal}}$, and coalescence phase $\varphi_{\rm{coal}}$. 
And then we denote those parameters as $\bm{\theta}=\{m, d, t_{\rm{coal}}, \varphi_{\rm{coal}}\}$.
In addition, the observed waveform is a linear combination of the two polarizations of the GWs, involving antenna patterns functions \cite{maggiore2007gravitational}, i.e. 
$\phi_{\rm{obs}}=F_{+}\phi_{+}+F_{\cross}\phi_{\cross}$. Parameters that determine the antenna pattern function, such as right ascension, declination and polarization angle are all set to $0$.
These settings are intended to simplify the simulation and do not significantly affect the results of this study.

In the following simulations, the threshold values introduced in Sec. \ref{Sec: Dismiss parameters} are set to $r_{\rm{th}}=5,~h_{\rm{th}}=5$. $r_{\rm{th}}$ is set to this value because $r\lesssim4$ for the PML and SIS models in the parameters we examine, as shown in Fig. \ref{Fig: O(1) coefficient}. 
For $h_{\rm{th}}$, as we have discussed in the last paragraph in Sec. \ref{Sec: Dismiss parameters}, it is set to be $h_{\rm{th}}=r_{\rm{th}}$.

The frequency range is fixed to $f_{\rm{min}}=1~\text{[Hz]}$ and $f_{\rm{max}}=1024~\text{[Hz]}$ unless otherwise noted. In order for our methodology to work, the frequency range must be broad enough
\footnote{
Our methodology would become even more powerful through multi-band observations. For instance, a GW150914-like signal, which is microlensed, could be observed both by a decihertz detector covering $10^{-2}\text{ Hz}$ to $1\text{ Hz}$ and by an XG ground-based detector covering $1\text{ Hz}$ to $1\text{ kHz}$. Applying the Kramers–Kronig relation across this extended bandwidth would reduce truncation errors and further enhance the effectiveness of the method.}
. We have accordingly chosen a frequency band similar to third generation ground based detectors \cite{Reitze:2019iox, punturo2010}.
The value of 1024 is used only for convenience
of numerical calculations. 
In higher or lower frequency bands, as mentioned in footnote\ref{footnote: mass range}, the target lens mass becomes lighter or heavier, respectively. Moreover, in Sec. \ref{Sec: ECC signal}, we investigate the frequency dependence of our methodology and find that it becomes increasingly effective in the lower frequency band for eccentric GWs.
Additionally, in this section, $K_{\rm{obs}}(f_{\rm{min}})$, $\Delta_{\rm{obs}}(f_{\rm{mid}})$, $|F_{\rm{obs}}(f_{\rm{min}})-1|$ and $|F_{\rm{obs}}(f_{\rm{mid}})-1|$ are calculated as averaged values over a neighborhood of $f_{\rm{min}}$ and $f_{\rm{mid}}$, since they may oscillate. 
Specifically, the range is chosen to be sufficiently narrow compared to $[f_{\rm{min}}, f_{\rm{max}}=1024f_{\rm{min}}]$, with the former being the average over $[f_{\rm{min}}, 1.7f_{\rm{min}}]$ and the latter over $[f_{\rm{mid}}/1.7, 1.7f_{\rm{mid}}]$.

\subsection{Cases where observed GWs are lensed}
\label{Sec: True signal}

\begin{figure}[t]
    \centering
    \includegraphics[scale=0.38]{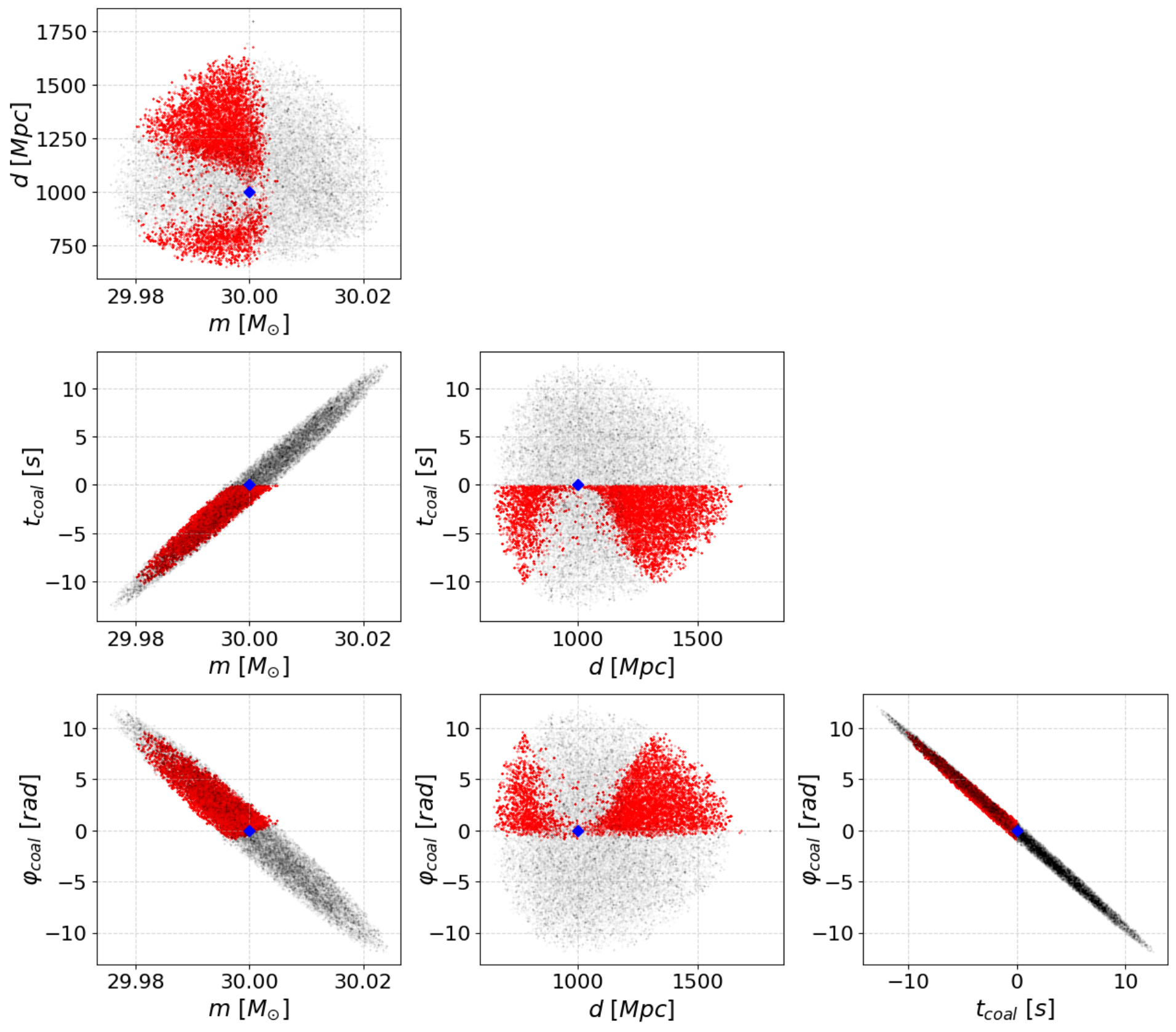}
    \caption{Simulation result in the case that the PML model is used for $F(f)$ in Eq. (\ref{phiobs GLsignal}). Parameters are $M_{Lz}=300M_{\odot},~y=0.1$. In the 4-dimensional parameter space of $\phi_{\rm{cir}}$, points are plotted uniformly and randomly in region $D$. Red points represent points satisfying $r\leq5$, and black points represent points satisfying $r>5$. Thus, red points cannot be dismissed and black points can be dismissed. The blue points represent the true parameters of $\phi_{\rm{cir}}$, $\bm{\theta}_T=\{30M_{\odot},~1000~\text{Mpc},~0~\text{s},~0~\text{rad}\}$.}
    \label{Fig: 4dim_PML}
\end{figure}

Here we examine the case where $\phi_{\rm{obs}}$ is a GL signal. That is,
\begin{equation}
    \phi_{\rm{obs}}(f)=F_{\rm{PML/SIS}}(f)\phi_{\rm{cir}}(f;\bm{\theta}_{T}) \label{phiobs GLsignal}
\end{equation}
with $F_{\rm{PML/SIS}}(f)$ calculated from PML and SIS lens models, where $\phi_{\rm{cir}}$ denotes the GWs of quasi-circular orbit and $\bm{\theta}_T=\{30M_{\odot},~1000~\text{Mpc},~0~\text{s},~0~\text{rad}\}$ are the true parameters.
By choosing a template waveform \footnote{We use the \textsc{TaylorF2} approximant (see, e.g., \cite{Buonanno:2009zt}) to simulate quasi-circular waveforms.} as $\phi_{\rm{cir}}(f;\bm{\theta})$, we obtain
\begin{equation}
    F_{\rm{obs}}(f;\bm{\theta})=F_{\rm{PML/SIS}}(f)\frac{\phi_{\rm{cir}}(f;\bm{\theta}_T)}{\phi_{{\rm{cir}}}(f;\bm{\theta})}
\end{equation}
as the observed amplification factor. Obviously, this is a false amplification factor when $\bm{\theta}$ deviates from $\bm{\theta}_T$. In the following, simulations are performed for PML and SIS models.

\subsubsection{Point mass lens}
\label{Sec: PML signal}

\begin{figure}[t]
    \centering
    \includegraphics[scale=0.38]{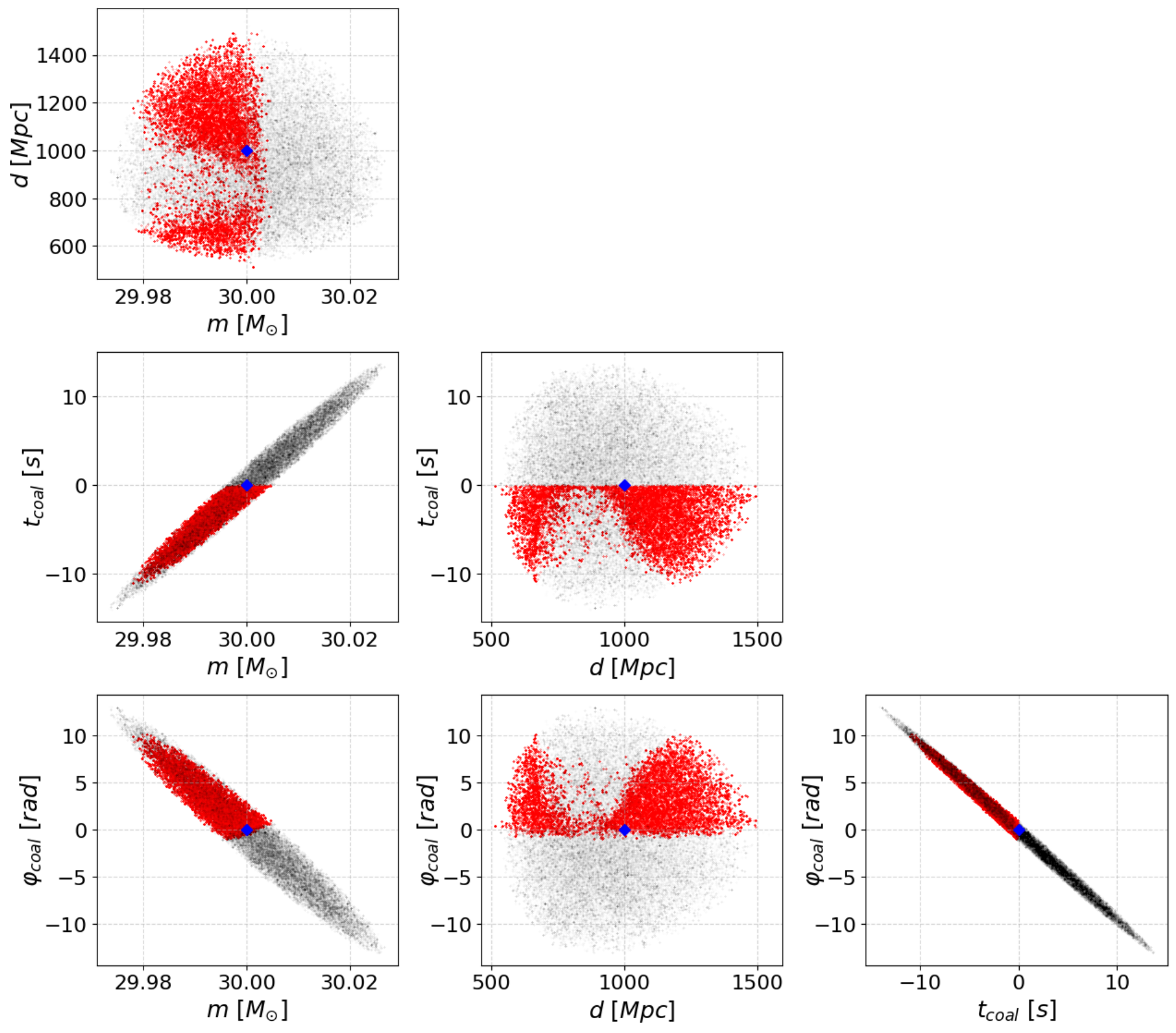}
    \caption{Simulation result in the case that the SIS model is used for $F(f)$ in Eq. (\ref{phiobs GLsignal}). Parameters are $M_{Lz}=300M_{\odot},~y=0.1$. Compared to the result for the PML model in Fig. \ref{Fig: 4dim_PML}, the distribution of points that can be dismissed (black points) and those that cannot (red points) is very similar.}
    \label{Fig: 4dim_SIS}
\end{figure}

Here, we use the PML model for the amplification factor. The parameters of lens are $M_{Lz}=300M_{\odot},~y=0.1$ and $F_{\rm{PML}}(f)$ for this case is shown in the left panel of Fig. \ref{Fig: ampfac_PML_SIS}. 
Fig. \ref{Fig: 4dim_PML} shows the result. This is the plot of the 4-dimensional parameter space of $\bm{\theta}$, with the points uniformly and randomly sampled within region $D$. The red points are those not dismissed by the KK relation, i.e., $r(\bm{\theta})\leq5$, and the black points are those dismissed, i.e., $r(\bm{\theta})>5$. The blue point is the true parameter $\bm{\theta}_T$. 
It is located on the boundary between the red and black points but is not dismissed as expected, indicating that our methodology is consistent. The reason why $\bm{\theta}_{T}$ is on the boundary can be explained as follows in terms of the parameter $t_{\rm{coal}}$: The KK relation is founded on the causality and $t_{\rm{coal}}$ is a parameter related to time shift, that is, the arrival time of the signal. Therefore, changing $t_{\rm{coal}}$ from its true value in the direction in which the signal arrives earlier is prohibited by the KK relation, but not in the opposite direction.
For $m$ and $\varphi_{\rm{coal}}$, since they are parameters that change the phase of the template waveform as well as $t_{\rm{coal}}$, they are considered to mimic the time shift in this simulation and exhibit the same behavior as $t_{\rm{coal}}$.
Thus, $\bm{\theta}_{T}$ is located at the boundary between the dismissed and non-dismissed regions.

This result also shows that the KK relation can restrict the parameter space. It should be noted, however, that the true parameters cannot be obtained from the KK relation alone.

\subsubsection{Singular isothermal sphere}
\label{Sec: SIS signal}

\begin{figure}[t]
    \centering
    \includegraphics[scale=0.6]{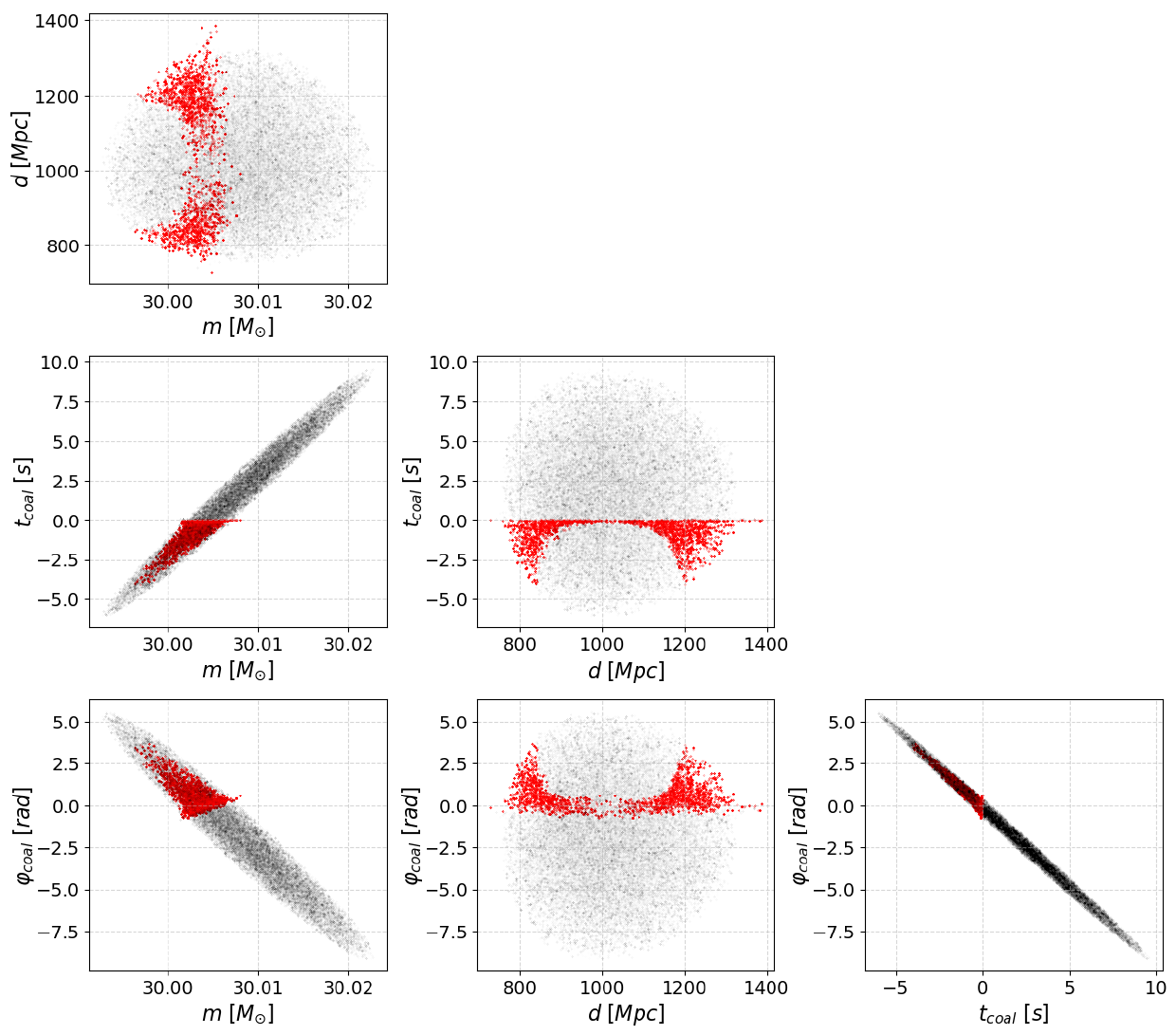}
    \caption{Simulation result in the case that $\phi_{\rm{obs}}=\phi_{\rm{ecc}}$. Parameters are $e=0.01$ and $\bm{\theta}_0=\{30M_{\odot},~1000~\text{Mpc},~0~\text{s},~0~\text{rad}\}$. In this case, $\phi_{\rm ecc}$ can not be dismissed as a non-GL signal since there are many points that can not be dismissed (red points).}
    \label{Fig: 4dim_e0.010}
\end{figure}

Here, we simulate the same as in Sec. \ref{Sec: PML signal}, using the SIS model for the amplification factor. The parameters of lens are $M_{Lz}=300M_{\odot},~y=0.1$ and $F_{\rm{SIS}}(f)$ for this case is shown in the right panel of Fig. \ref{Fig: ampfac_PML_SIS}. The result shown in Fig. \ref{Fig: 4dim_SIS} is the same as for the PML model; in the SIS model, the KK relation can be used to dismiss parameters and restrict the parameter space of the template. It should be noted that, as in PML, it is not possible to obtain the true parameters from the KK relation alone.

\subsection{Cases where observed GWs are not lensed}
\label{Sec: False signal}

Here we examine the case where $\phi_{\rm{obs}}$ is not a GL signal. 
As such signals, we consider two GW sources: i) binary with eccentric orbit $\phi_{\rm{ecc}}$ and ii) binary in which one of the stars has its own angular momentum (spin) $\phi_{\rm{spin}}$. The deviation of these waveforms from $\phi_{\rm{cir}}$ is observed as the false amplification factor:
\begin{equation}
    F_{\rm{obs}}(f;\bm{\theta})=\frac{\phi_{\rm{ecc/spin}}(f; e, \bm{s}, \bm{\theta}_0)}{\phi_{\rm{cir}}(f;\bm{\theta})}~,\label{ampfac nonGL}
\end{equation}
where $e$ is the eccentricity of orbit, $\bm{s}$ is the spin of the star and $\bm{\theta}_0$ is the set of four parameters same as for $\phi_{\rm{cir}}$ and is fixed.
Since $\phi_{\rm{ecc/spin}}$ are not GL signals, 
$F_{\rm{obs}}$ is false for any $\bm{\theta}$. In the following, we examine how many parameters can be dismissed by Criterion (\ref{statement: r>O(1)}) and whether we can dismiss the non-GL signal $\phi_{\rm{ecc/spin}}$.

\subsubsection{GWs from eccentric orbit}
\label{Sec: ECC signal}

\begin{figure}[t]
    \centering
    \includegraphics[scale=0.45]{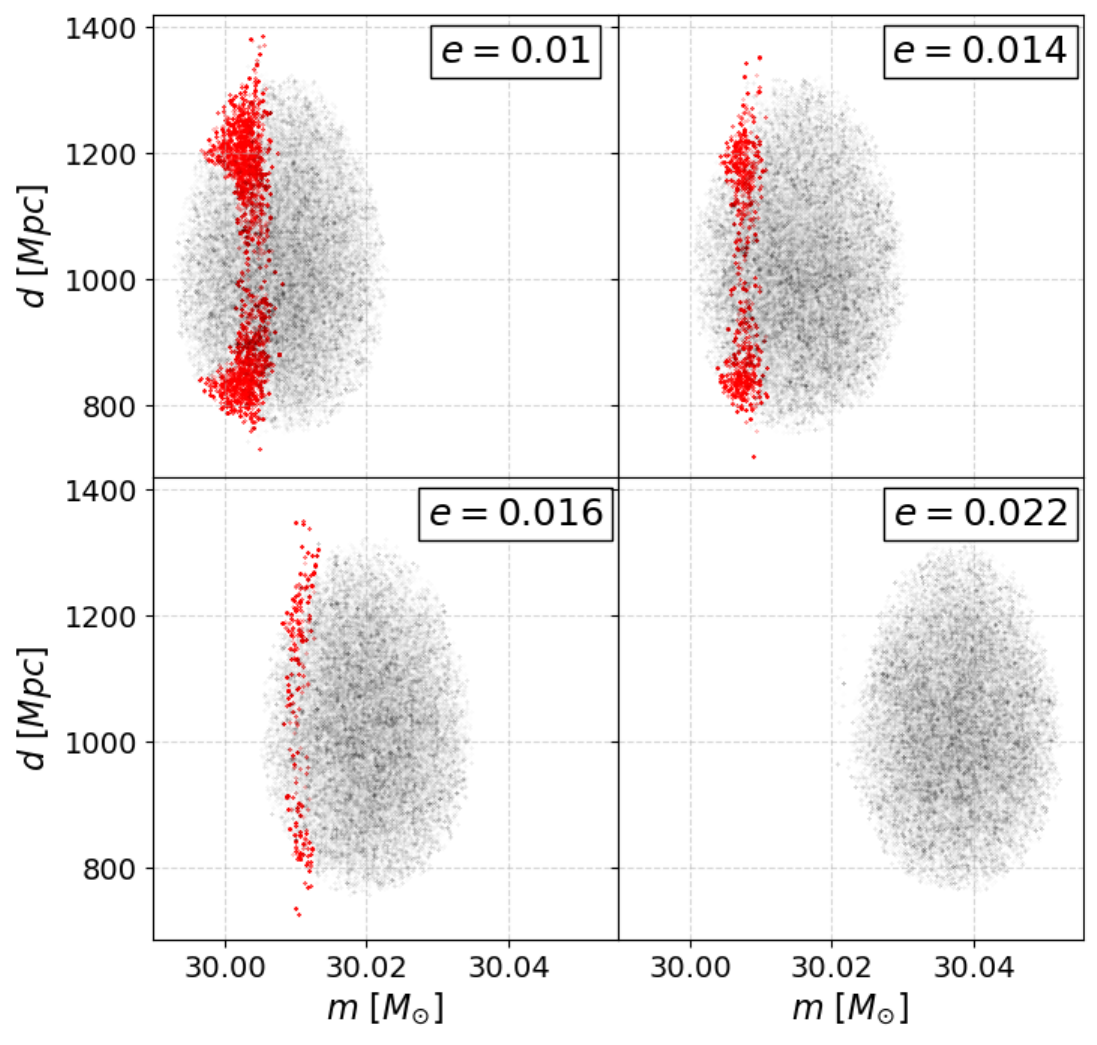}
    \caption{$m$-$d$ plots in Fig. \ref{Fig: 4dim_e0.010} for different eccentricities. The number of the points that can not be dismissed (red points) decreases as $e$ increases, eventually reaching zero at $e=0.022$. Region $D$, the parameter region where the amplification factor obtained from Eq. (\ref{ampfac nonGL}) behaves as if the low-frequency side is in the WO regime, moving away from the parameters of $\phi_{\rm cir}$, $\{30M_{\odot},~1000~\text{Mpc}\}$, as $e$ increases and the deviation of $\phi_{\rm ecc}$ from $\phi_{\rm cir}$ also increases.}
    \label{Fig: tileplot_ecc}
\end{figure}

\begin{figure}[t]
    \centering
    \begin{minipage}{0.4\columnwidth}
        \centering
        \includegraphics[width=1\columnwidth]{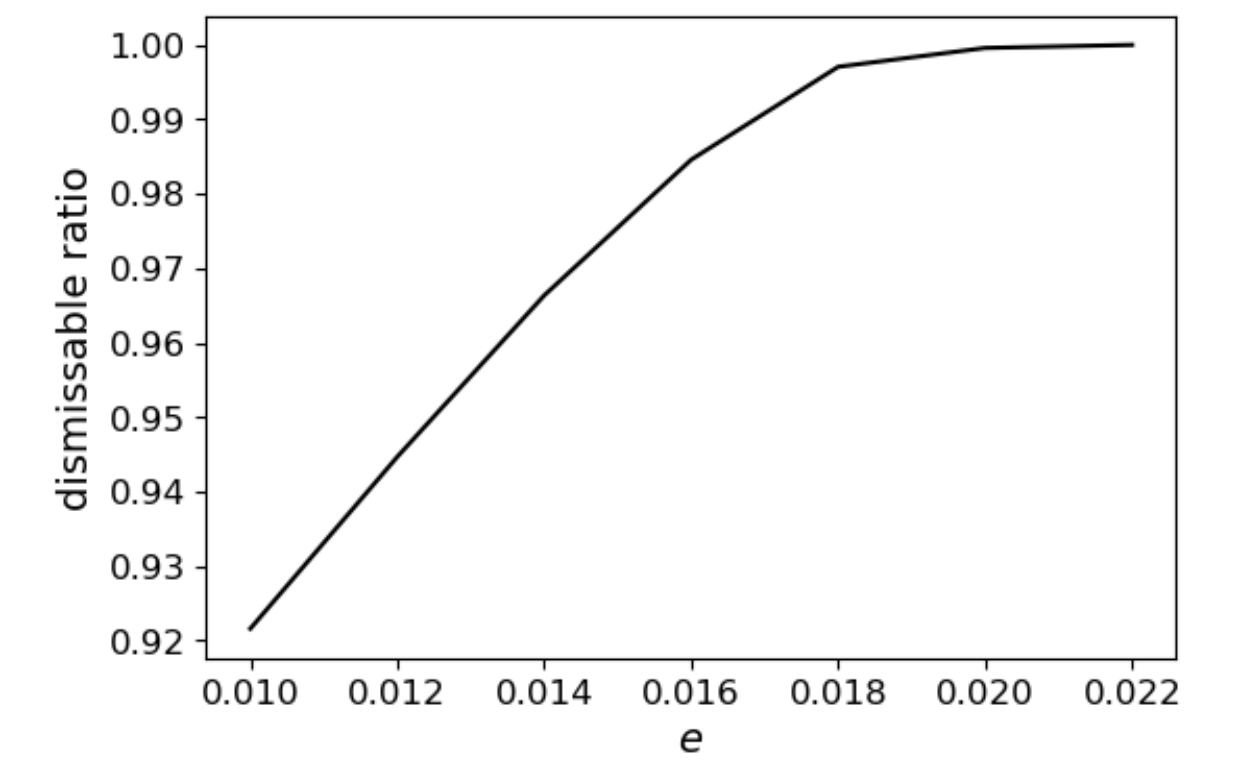}
    \end{minipage}
    \begin{minipage}{0.4\columnwidth}
        \centering
        \includegraphics[width=1\columnwidth]{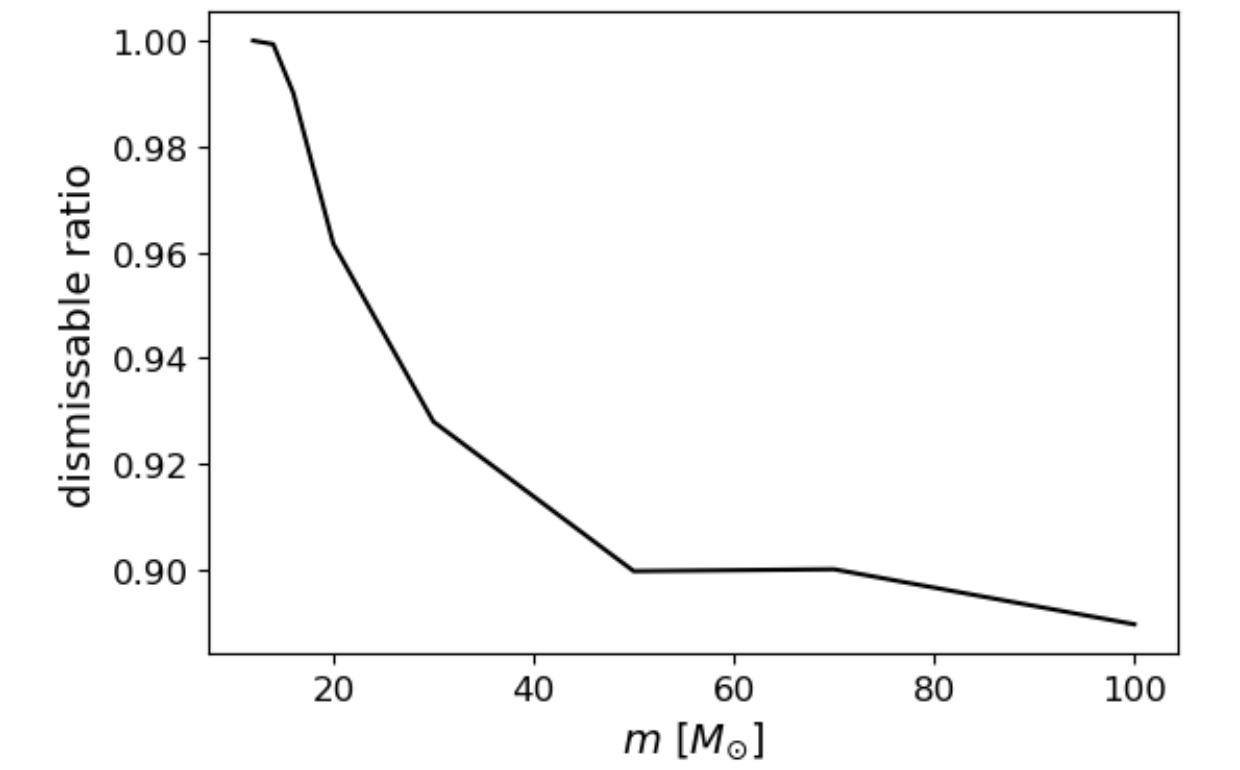}
    \end{minipage}
    
    \begin{minipage}{0.4\columnwidth}
        \centering
        \includegraphics[width=1\columnwidth]{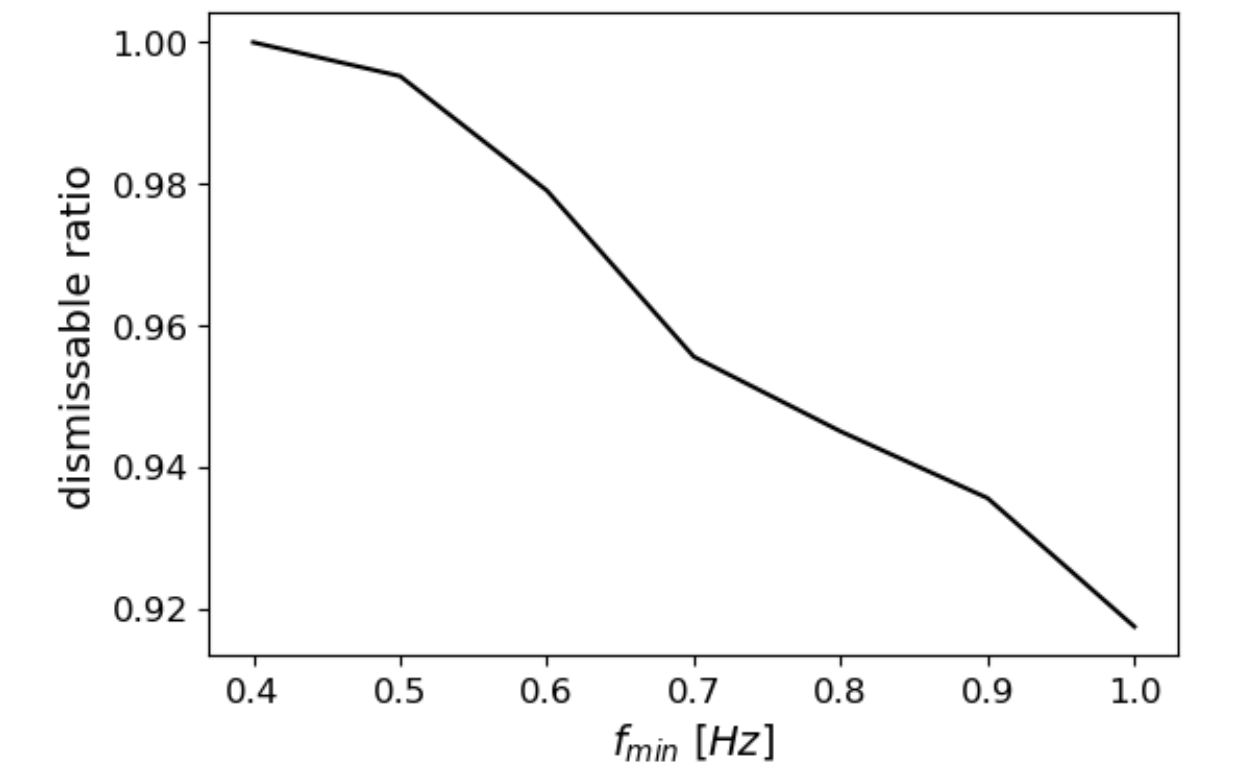}
    \end{minipage}
    \begin{minipage}{0.4\columnwidth}
        \centering
        \includegraphics[width=1\columnwidth]{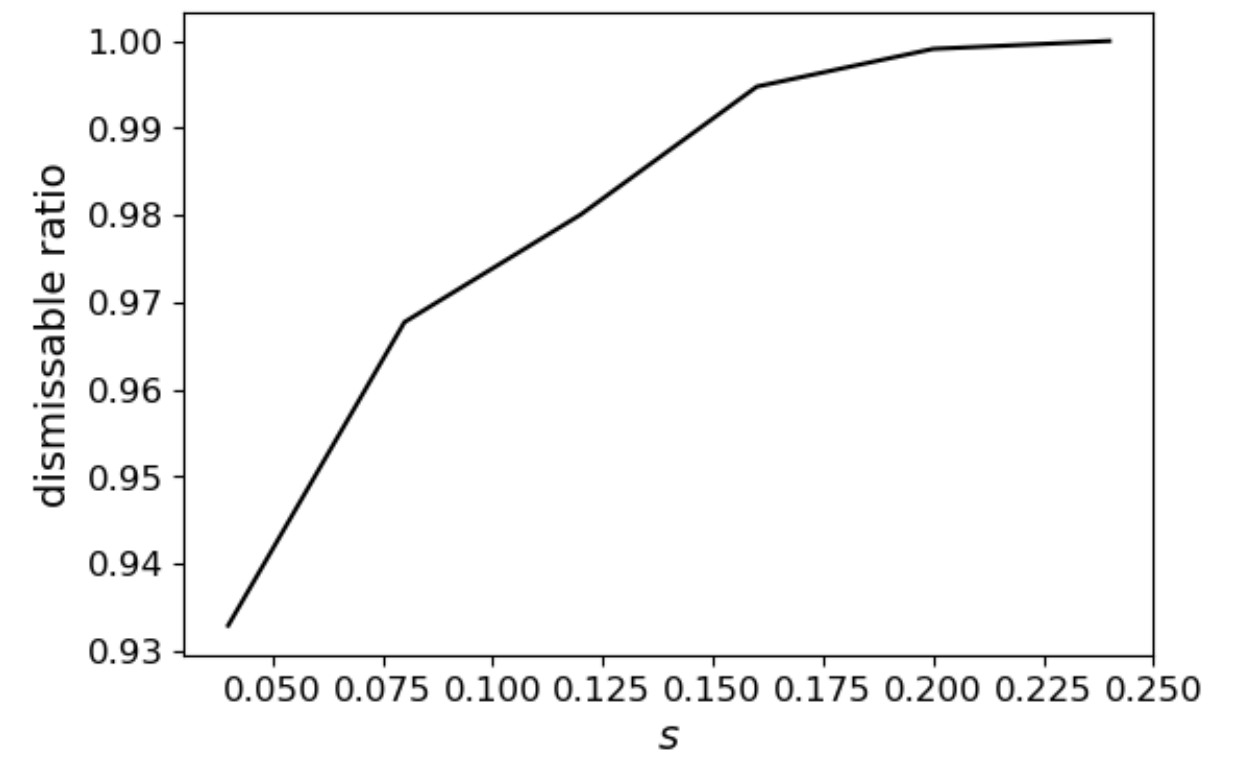}
    \end{minipage}
    
    \caption{Plots of dismissable ratio (the number of dismissable points divided by the total number of points) within the parameter region $D$ when the observed waveform is the non-GL signal. The top left, top right, and bottom left panels are for $\phi_{\rm{obs}}=\phi_{\rm{ecc}}$, and the bottom right panel is for $\phi_{\rm{obs}}=\phi_{\rm{spin}}$.}
    \label{Fig: dismissable ratio}
\end{figure}

Here we investigate the case $\phi_{\rm{obs}}=\phi_{\rm{ecc}}$ \footnote{We use the \textsc{TaylorF2Ecc} approximant to simulate eccentric waveforms \cite{Moore:2016qxz}.}. Fig. \ref{Fig: 4dim_e0.010} shows the result when $e = 0.01,~\bm{\theta}_0=\{30M_{\odot},~1000~\text{Mpc},~0~\text{s},~0~\text{rad}\}$. 
As it can be seen from Fig. \ref{Fig: 4dim_e0.010}, there are many points that cannot be dismissed (i.e., red points) in this case. In other words, $\phi_{\rm{ecc}}(e=0.01)$ mimics a GL signal in the context of the KK relation for some templates, and we cannot dismiss such GWs as non-lensed GWs at least for the frequency range we adopt. 

Next, we investigate how the results change by varying the eccentricity. We show the results in Fig. \ref{Fig: tileplot_ecc} and in the top left panel of Fig. \ref{Fig: dismissable ratio}. It can be seen that as $e$ increases, the ratio of dismissable points increases, and eventually all points can be dismissed. This means that we can dismiss $\phi_{\rm{ecc}}(e\geq0.022)$ as not a GL signal with WO.
This is because the larger $e$, the greater the deviation of $\phi_{\rm{ecc}}$ from $\phi_{\rm{cir}}$, and the greater the deviation of $F_{\rm{obs}}$ from the physical amplification factor, and thus the easier it is to dismiss.

We also examine the dependence on the binary mass $m$. Here, $e=0.01,~\bm{\theta}_0=\{m,~1000~\text{Mpc},~0~\text{s},~0~\text{rad}\}$ and the result for varying $m$ is shown in the top right panel of Fig. \ref{Fig: dismissable ratio}. It shows that the ratio of dismissable points increases as $m$ is smaller, which can be explained as follows: when $m$ is small, the number of cycles of the binary stars in the same frequency range increases. Therefore, the deviation from the circular orbit due to the eccentricity that accumulates with each cycle also increases, and $\phi_{\rm{ecc}}$ is more easily dismissed.

Furthermore, we show the dependence on $f_{\rm{min}}$. Here the ratio of $f_{\rm{min}}$ to $f_{\rm{max}}$ is fixed as $f_{\rm{max}}=1024f_{\rm{min}}$, and other parameters are $e=0.01,~\bm{\theta}_0=\{30M_{\odot},~1000~\text{Mpc},~0~\text{s},~0~\text{rad}\}$. 
The bottom left panel of Fig. \ref{Fig: dismissable ratio} is the result.
We find that the dismissable ratio increases as $f_{\rm min}$ is decreased.
This can be interpreted as follows: the orbit of binary stars is initially elliptical but approaches a circular orbit as it loses energy by emitting GWs. Therefore, the effect of the eccentricity on the waveform increases at lower frequencies, and $\phi_{\rm{ecc}}$ becomes more easily dismissable.

\begin{figure}[t]
    \centering
    \includegraphics[scale=0.6]{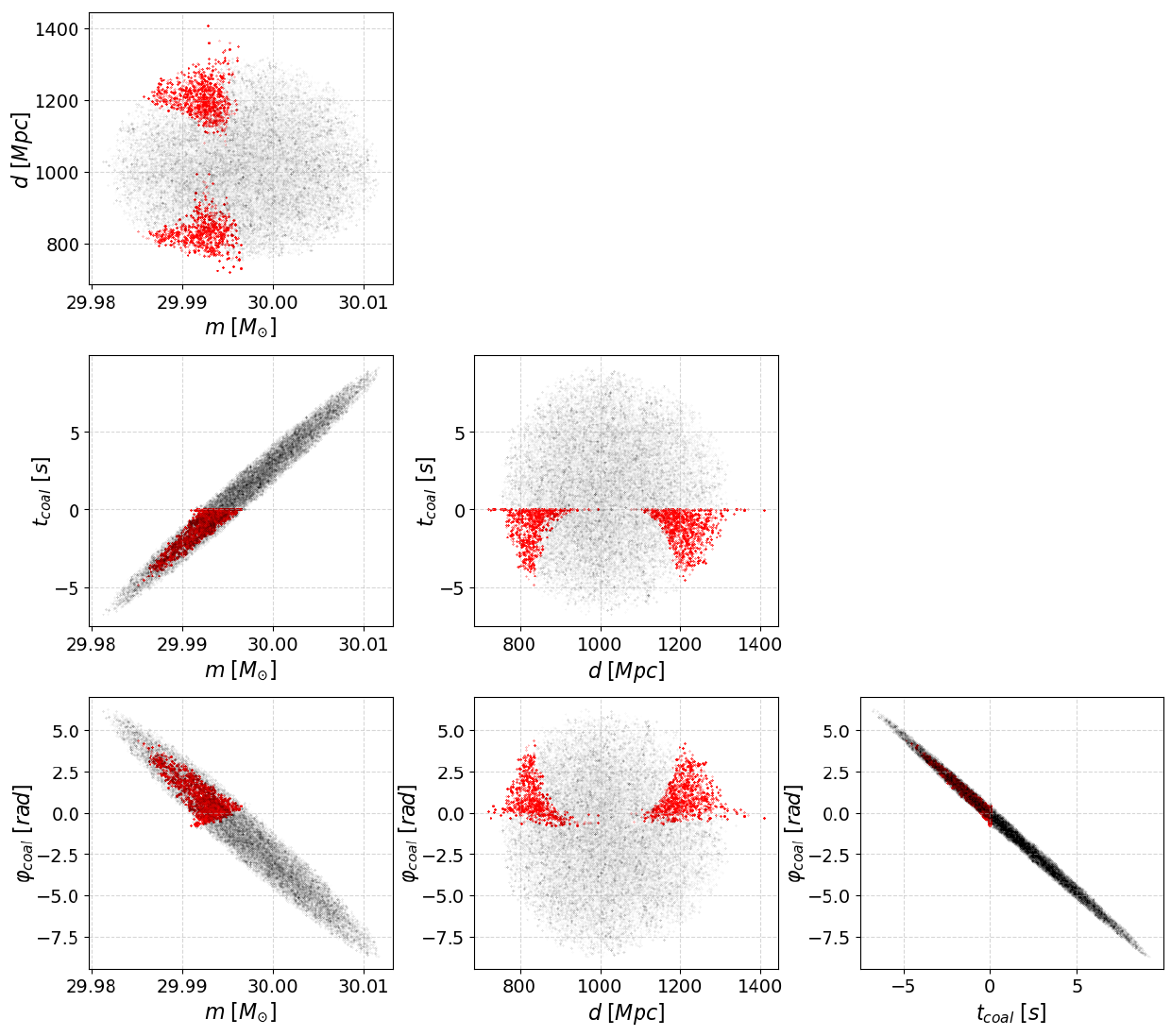}
    \caption{Simulation result in the case that $\phi_{\rm{obs}}=\phi_{\rm{spin}}$. Parameters are $s=0.04$ and $\bm{\theta}_0=\{30M_{\odot},~1000~\text{Mpc},~0~\text{s},~0~\text{rad}\}$. Same as the result for $\phi_{\rm ecc}(e=0.01)$ in Fig. \ref{Fig: 4dim_e0.010}, in this case, $\phi_{\rm spin}$ can not be dismissed as a non-GL signal since there are many points that can not be dismissed (red points).}
    \label{Fig: 4dim_s0.04}
\end{figure}

\subsubsection{GWs from spinning binary}
\label{Sec: SPIN signal}

\begin{figure}[t]
    \centering
    \includegraphics[scale=0.45]{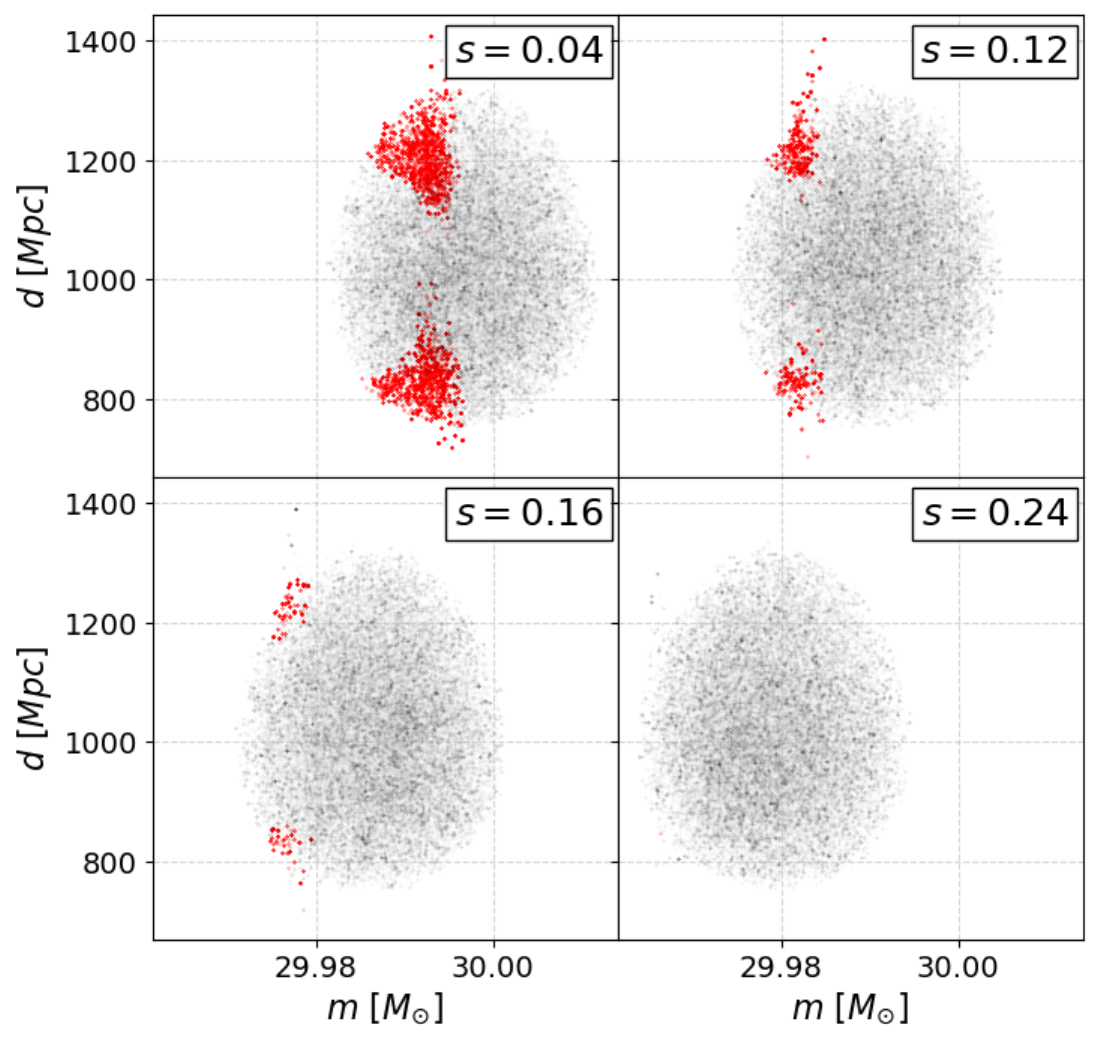}
    \caption{$m$-$d$ plots in Fig. \ref{Fig: 4dim_s0.04} for different spins. The number of the points that can not be dismissed (red points) decreases as $s$ increases, eventually reaching zero at $s=0.24$. Region $D$, the parameter region where the amplification factor obtained from Eq. (\ref{ampfac nonGL}) behaves as if the low-frequency side is in the WO regime, moving away from the parameters of $\phi_{\rm cir}$, $\{30M_{\odot},~1000~\text{Mpc}\}$, as $s$ increases and the deviation of $\phi_{\rm spin}$ from $\phi_{\rm cir}$ also increases.}
    \label{Fig: tileplot_spin}
\end{figure}

We also simulate in the case that the observed waveform is GWs with one of the binary stars having spin, i.e. $\phi_{\rm{obs}}=\phi_{\rm{spin}}$ \footnote{We use the \textsc{IMRPhenomPv3} approximant to simulate spinning waveforms \cite{Khan:2018fmp}.}. 
We assume that the spin is perpendicular to the plane of rotation of the binary stars. That is, if we take that plane to be the xy-plane, then $\bm{s}=\{0,0,s\}$ where $s$ is the dimensionless spin\footnote{$s=J/Gm^2$, where $J$ is the angular momentum of the spin.}. 
We examined for several values of $s$ while we set the other parameters to be $\bm{\theta}_0=\{30M_{\odot},~1000~\text{Mpc},~0~\text{s},~0~\text{rad}\}$. Fig. \ref{Fig: 4dim_s0.04} shows the result when $s=0.04$. 
In this case, there are many points that cannot be dismissed (i.e., red points) and this $\phi_{\rm{spin}}$ is not dismissable. 
However, as can be seen from Fig. \ref{Fig: tileplot_spin} and the bottom right panel of Fig. \ref{Fig: dismissable ratio}, as $s$ is increased, the proportion of points that can be dismissed increases, and eventually $\phi_{\rm{spin}}$ can be dismissed as a non-GL signal. This behavior is similar to the case of GWs with the eccentricity and can be understood that the larger $s$ is, the larger the deviation of $\phi_{\rm{spin}}$ from $\phi_{\rm{cir}}$ becomes and easier it is to dismiss.

\section{Conclusion}
\label{Sec: Conclusion}

We provided a proof-of-principle study of the application of KK-relation for a model-independent identification of gravitationally microlensed GW events exhibiting wave-optics effects. Under the idealized assumption of no noise within a finite frequency band (and infinite noise outside it), the observed waveform is divided by a template quasi-circular, non-spinning waveform, to acquire the observed amplification factor. If the unlensed template waveform does not match the true source waveform, or if the observed waveform is unlensed to begin with, a false amplification factor will be obtained if the observed waveform exhibits any deviation from quasi-circularity (and spin). Then, we proposed the methodology for dismissing false amplification factors by exploiting the violation of the KK relation. 
If the actual violation exceeds the estimation in Eq. (\ref{estimation obs}), the KK relation is considered to be violated and the amplification factor is dismissed. The indicator of this is $r$ defined by Eq. (\ref{r}), and we dismiss the observed amplification factor when $r>r_{\rm{th}}$, as in the statement \ref{statement: r>O(1)}.

We also examined to what extent our methodology can dismiss false amplification factors.
As a result, in cases where the observed waveform is a GL signal, some parameters could be dismissed and the parameter space was restricted. In addition, the true parameter was not dismissed, confirming that the KK relation works consistently.
On the other hand, when the observed waveform is not a GL signal, it was found that not only can the parameter space be restricted, but also the observed waveform, in some cases, can be dismissed as not a GL signal with WO. We considered GWs with eccentricity and with spin as the observed waveforms, and found that in both cases, the observed waveforms can be dismissed if the eccentricity or spin is sufficiently large.

This study demonstrates the applicability of the KK relation: it can restrict the parameter space of the template in GL observations and can dismiss those that are not GL signals with WO. We therefore believe that KK relation could potentially be used as a powerful analytical tool in future observations, for signals with large signal-to-noise ratios.

It should be noted, however, that this study ignores any detector noise. In actual observations, there will be errors in the obtained amplification factors due to noise, and how to use the KK relation after taking this into account remains a non-trivial issue that we leave for future work.


\appendix
\section{$\mathcal{O}(1)$ coefficient in Eq. (\ref{estimation})}
\label{Sec: Appendix}

\begin{figure}[t]
    \centering
    \includegraphics[scale=0.8]{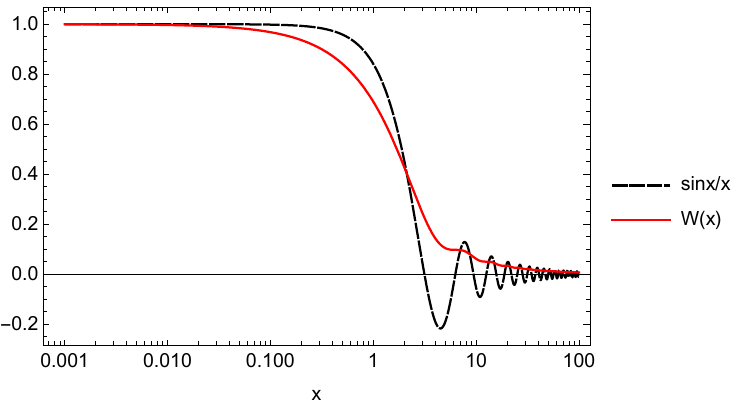}
    \caption{Comparison of $\mathrm{sinc}(x)$ and $W(x)$. The orders of magnitude coincide in all x, but $\mathrm{sinc}(x)$ oscillates while $W(x)$ does not.}
    \label{Fig: filterfunctions}
\end{figure}

In the Born approximation and thin-lens approximation, and for the symmetric lenses, $K(\omega_{\rm min})$ and $\Delta_{\rm tr}(\omega_{\rm mid})$ with the condition $\omega_{\rm min}\ll\omega_{\rm mid}\ll\omega_{\rm max}$ are given by\cite{Tanaka:2023mvy}
\begin{eqnarray}
    K(\omega_{\rm min})
    &=&\int\frac{d^2k_\perp}{(2\pi)^2}\tilde{\kappa}(k_\perp)e^{i\bm{k}_\perp\cdot\bar{\eta}}~\mathrm{sinc}\qty(\frac{r_F^2(\omega_{\rm min})k_\perp^2}{2})~,\label{Appendix: K}\\
    \Delta_{\rm tr}(\omega_{\rm mid})
    &\simeq&\int\frac{d^2k_\perp}{(2\pi)^2}\tilde{\kappa}(k_\perp)e^{i\bm{k}_\perp\cdot\bar{\eta}}~W\qty(\frac{r_F^2(\omega_{\rm min})k_\perp^2}{2})~,\label{Appendix: Delta_tr}
\end{eqnarray}
where
\begin{eqnarray}
    \mathrm{sinc}(x)
    &\equiv&\frac{\sin x}{x}~,\\
    W(x)
    &\equiv&-\frac{2}{\pi}\qty[\frac{\cos x -1}{x}-\int_x^\infty \frac{\sin t}{t} dt]~,
\end{eqnarray}
$r_F$ is the Fresnel scale defined by
\begin{equation}
    r_F(\omega)\equiv\sqrt{\frac{D_LD_{LS}}{\omega D_S}}~,
\end{equation}
$\tilde{\kappa}(\bm{k}_\perp)$ is the Fourier transform of dimensionless surface mass density and $\bar{\eta}=\eta D_L/D_S$.
As shown in \cite{Tanaka:2023mvy}, the estimation (\ref{estimation}) is based on the similarity in the shape of $\mathrm{sinc}(x)$ and $W(x)$: the orders of their magnitude coincide in all $x$ (see Fig. \ref{Fig: filterfunctions}).
The only difference is that $\mathrm{sinc}(x)$ oscillates while $W(x)$ does not, resulting in the difference between $K(\omega_{\rm min})$ and $\Delta_{\rm tr}(\omega_{\rm mid})$ through the integration in Eq. (\ref{Appendix: K}) and (\ref{Appendix: Delta_tr}).

For example, in the PML model, $\tilde{\kappa}(k_\perp)=const.$, and $k_\perp<1/\bar{\eta}$ mainly contributes to the integral due to the exponential factors in (\ref{Appendix: K}) and (\ref{Appendix: Delta_tr}). Thus, as $\bar{\eta}$ decreases, the effective integration range expands, leading to a greater increase in $\Delta_{\rm tr}(\omega_{\rm mid})$ relative to $K(\omega_{\rm min})$. Consequently, when $M_{Lz}$ decreases while keeping $y$ fixed, $\bar{\eta}$ also decreases, resulting in an increase in $|\Delta_{\rm tr}(\omega_{\rm mid})|/|K(\omega_{\rm min})|$ as shown in Fig. \ref{Fig: O(1) coefficient}.

In other lens models, $\tilde{\kappa}(k_\perp)$ is suppressed at large $k_\perp$: for example, in the SIS model, $\tilde{\kappa}(k_\perp)\propto k_\perp^{-1}$. Therefore, the effective integral range in Eq. (\ref{Appendix: K}) and (\ref{Appendix: Delta_tr}) is narrower than in the PML model, and $|\Delta_{\rm tr}(\omega_{\rm mid})|/|K(\omega_{\rm min})|$ is expected to be smaller than in the PML model.

\section*{Acknowledgements}
This work was supported by JSPS KAKENHI Grant Number 24KJ1094 (ST) and by JSPS KAKENHI Grant Number JP23K03411 (TS). S.J.K. gratefully acknowledges support from SERB grants SRG/2023/000419 and MTR/2023/000086.

\textit{Software}: \texttt{PyCBC} \cite{pycbc}, \texttt{NumPy} \cite{vanderWalt:2011bqk}, \texttt{SciPy} \cite{Virtanen:2019joe}, \texttt{astropy} \cite{2013A&A...558A..33A, 2018AJ....156..123A}, \texttt{Matplotlib} \cite{Hunter:2007}, \texttt{jupyter} \cite{jupyter}.

\bibliography{ref}

@book{misner2017gravitation,
  title={Gravitation},
  author={Misner, C.W. and Thorne, K.S. and Wheeler, J.A. and Kaiser, D.I.},
  isbn={9781400889099},
  url={https://books.google.co.jp/books?id=zAAuDwAAQBAJ},
  year={2017},
  publisher={Princeton University Press}
}

@article{Bailes:2021tot,
    author = "Bailes, M. and others",
    title = "{Gravitational-wave physics and astronomy in the 2020s and 2030s}",
    doi = "10.1038/s42254-021-00303-8",
    journal = "Nature Rev. Phys.",
    volume = "3",
    number = "5",
    pages = "344--366",
    year = "2021"
}

@article{peters1974index,
  title={Index of refraction for scalar, electromagnetic, and gravitational waves in weak gravitational fields},
  author={Peters, Philip C},
  journal={Physical Review D},
  volume={9},
  number={8},
  pages={2207},
  year={1974},
  publisher={APS}
}

@article{Bartelmann:2010fz,
    author = "Bartelmann, Matthias",
    title = "{Gravitational Lensing}",
    eprint = "1010.3829",
    archivePrefix = "arXiv",
    primaryClass = "astro-ph.CO",
    doi = "10.1088/0264-9381/27/23/233001",
    journal = "Class. Quant. Grav.",
    volume = "27",
    pages = "233001",
    year = "2010"
}

@article{Nakamura:1999uwi,
    author = "Nakamura, Takahiro T. and Deguchi, Shuji",
    title = "{Wave Optics in Gravitational Lensing}",
    doi = "10.1143/ptps.133.137",
    journal = "Prog. Theor. Phys. Suppl.",
    volume = "133",
    pages = "137--153",
    year = "1999"
}

@article{ruffa1999gravitational,
  title={Gravitational lensing of gravitational waves},
  author={Ruffa, Anthony A},
  journal={The Astrophysical Journal},
  volume={517},
  number={1},
  pages={L31},
  year={1999},
  publisher={IOP Publishing}
}

@article{Takahashi:2003ix,
    author = "Takahashi, Ryuichi and Nakamura, Takashi",
    title = "{Wave effects in gravitational lensing of gravitational waves from chirping binaries}",
    eprint = "astro-ph/0305055",
    archivePrefix = "arXiv",
    doi = "10.1086/377430",
    journal = "Astrophys. J.",
    volume = "595",
    pages = "1039--1051",
    year = "2003"
}

@article{Suyama:2020lbf,
    author = "Suyama, Teruaki",
    title = "{On arrival time difference between lensed gravitational waves and light}",
    eprint = "2003.11748",
    archivePrefix = "arXiv",
    primaryClass = "gr-qc",
    doi = "10.3847/1538-4357/ab8d3f",
    journal = "Astrophys. J.",
    volume = "896",
    number = "1",
    pages = "46",
    year = "2020"
}

@book{lucarini2005kramers,
  title={Kramers-Kronig Relations in Optical Materials Research},
  author={Lucarini, V. and Saarinen, J.J. and Peiponen, K.E. and Vartiainen, E.M.},
  isbn={9783540236733},
  lccn={2005921306},
  series={Springer Series in Optical Sciences},
  url={https://books.google.co.jp/books?id=U21PMIXyk7IC},
  year={2005},
  publisher={Springer Berlin Heidelberg}
}

@article{Suyama:2005mx,
    author = "Suyama, Teruaki and Takahashi, Ryuichi and Michikoshi, Shugo",
    title = "{Wave propagation in a weak gravitational field and the validity of the thin lens approximation}",
    eprint = "astro-ph/0505023",
    archivePrefix = "arXiv",
    reportNumber = "KUNS-1969",
    doi = "10.1103/PhysRevD.72.043001",
    journal = "Phys. Rev. D",
    volume = "72",
    pages = "043001",
    year = "2005"
}

@article{Mizuno:2022xxp,
    author = "Mizuno, Morifumi and Suyama, Teruaki",
    title = "{Weak lensing of gravitational waves in wave optics: Beyond the Born approximation}",
    eprint = "2210.02062",
    archivePrefix = "arXiv",
    primaryClass = "astro-ph.CO",
    month = "10",
    year = "2022"
}

@article{Takahashi:2005ug,
    author = "Takahashi, Ryuichi",
    title = "{Amplitude and phase fluctuations for gravitational waves propagating through inhomogeneous mass distribution in the universe}",
    eprint = "astro-ph/0511517",
    archivePrefix = "arXiv",
    doi = "10.1086/503323",
    journal = "Astrophys. J.",
    volume = "644",
    pages = "80--85",
    year = "2006"
}

@book{maggiore2007gravitational,
  title={Gravitational waves: Volume 1: Theory and experiments},
  author={Maggiore, Michele},
  year={2007},
  publisher={OUP Oxford}
}

@book{schneider1999gravitational,
  title={Gravitational Lenses},
  author={Schneider, P. and Ehlers, J. and Falco, E.E.},
  isbn={9783540665069},
  lccn={99049728},
  series={Astronomy and Astrophysics Library},
  url={https://books.google.co.jp/books?id=sPAIgy9QGBsC},
  year={1999},
  publisher={Springer}
}

@article{Oguri:2020ldf,
    author = "Oguri, Masamune and Takahashi, Ryuichi",
    title = "{Probing Dark Low-mass Halos and Primordial Black Holes with Frequency-dependent Gravitational Lensing Dispersions of Gravitational Waves}",
    eprint = "2007.01936",
    archivePrefix = "arXiv",
    primaryClass = "astro-ph.CO",
    doi = "10.3847/1538-4357/abafab",
    journal = "Astrophys. J.",
    volume = "901",
    number = "1",
    pages = "58",
    year = "2020"
}

@article{Babak:2021mhe,
    author = "Babak, Stanislav and Petiteau, Antoine and Hewitson, Martin",
    title = "{LISA Sensitivity and SNR Calculations}",
    eprint = "2108.01167",
    archivePrefix = "arXiv",
    primaryClass = "astro-ph.IM",
    reportNumber = "LISA-LCST-SGS-TN-001",
    month = "8",
    year = "2021"
}

@article{Tambalo:2022wlm,
    author = "Tambalo, Giovanni and Zumalac\'arregui, Miguel and Dai, Liang and Cheung, Mark Ho-Yeuk",
    title = "{Gravitational wave lensing as a probe of halo properties and dark matter}",
    eprint = "2212.11960",
    archivePrefix = "arXiv",
    primaryClass = "astro-ph.CO",
    doi = "10.1103/PhysRevD.108.103529",
    journal = "Phys. Rev. D",
    volume = "108",
    number = "10",
    pages = "103529",
    year = "2023"
}

@article{Savastano:2023spl,
    author = "Savastano, Stefano and Tambalo, Giovanni and Villarrubia-Rojo, Hector and Zumalacarregui, Miguel",
    title = "{Weakly lensed gravitational waves: Probing cosmic structures with wave-optics features}",
    eprint = "2306.05282",
    archivePrefix = "arXiv",
    primaryClass = "gr-qc",
    doi = "10.1103/PhysRevD.108.103532",
    journal = "Phys. Rev. D",
    volume = "108",
    number = "10",
    pages = "103532",
    year = "2023"
}

@article{Tanaka:2023mvy,
    author = "Tanaka, So and Suyama, Teruaki",
    title = "{Kramers-Kronig relation in gravitational lensing}",
    eprint = "2303.05650",
    archivePrefix = "arXiv",
    primaryClass = "gr-qc",
    doi = "10.1103/PhysRevD.108.044015",
    journal = "Phys. Rev. D",
    volume = "108",
    number = "4",
    pages = "044015",
    year = "2023"
}

@article{LIGOScientific:2014pky,
    author = "Aasi, J. and others",
    collaboration = "LIGO Scientific",
    title = "{Advanced LIGO}",
    eprint = "1411.4547",
    archivePrefix = "arXiv",
    primaryClass = "gr-qc",
    doi = "10.1088/0264-9381/32/7/074001",
    journal = "Class. Quant. Grav.",
    volume = "32",
    pages = "074001",
    year = "2015"
}

@article{Tambalo:2022plm,
    author = "Tambalo, Giovanni and Zumalac\'arregui, Miguel and Dai, Liang and Cheung, Mark Ho-Yeuk",
    title = "{Lensing of gravitational waves: efficient wave-optics methods and validation with symmetric lenses}",
    eprint = "2210.05658",
    archivePrefix = "arXiv",
    primaryClass = "gr-qc",
    month = "10",
    year = "2022"
}

@article{Nakamura:1997sw,
    author = "Nakamura, Takahiro T.",
    title = "{Gravitational lensing of gravitational waves from inspiraling binaries by a point mass lens}",
    reportNumber = "UTAP-272-97, YITP-97-61",
    doi = "10.1103/PhysRevLett.80.1138",
    journal = "Phys. Rev. Lett.",
    volume = "80",
    pages = "1138--1141",
    year = "1998"
}

@article{TheVirgo:2014hva,
      author         = "Acernese, F. and others",
      title          = "{Advanced Virgo: a second-generation interferometric
                        gravitational wave detector}",
      collaboration  = "Virgo",
      journal        = "Class. Quant. Grav.",
      volume         = "32",
      year           = "2015",
      number         = "2",
      pages          = "024001",
      doi            = "10.1088/0264-9381/32/2/024001",
      eprint         = "1408.3978",
      archivePrefix  = "arXiv",
      primaryClass   = "gr-qc",
      SLACcitation   = "%%CITATION = ARXIV:1408.3978;%%"
}

@article{LIGOScientific:2021izm,
    author = "Abbott, R. and others",
    collaboration = "LIGO Scientific, VIRGO",
    title = "{Search for Lensing Signatures in the Gravitational-Wave Observations from the First Half of LIGO\textendash{}Virgo\textquoteright{}s Third Observing Run}",
    eprint = "2105.06384",
    archivePrefix = "arXiv",
    primaryClass = "gr-qc",
    reportNumber = "LIGO-P2000400",
    doi = "10.3847/1538-4357/ac23db",
    journal = "Astrophys. J.",
    volume = "923",
    number = "1",
    pages = "14",
    year = "2021"
}

@article{LIGOScientific:2023bwz,
    author = "Abbott, R. and others",
    collaboration = "LIGO Scientific, KAGRA, VIRGO",
    title = "{Search for Gravitational-lensing Signatures in the Full Third Observing Run of the LIGO\textendash{}Virgo Network}",
    eprint = "2304.08393",
    archivePrefix = "arXiv",
    primaryClass = "gr-qc",
    reportNumber = "LIGO-P2200031",
    doi = "10.3847/1538-4357/ad3e83",
    journal = "Astrophys. J.",
    volume = "970",
    number = "2",
    pages = "191",
    year = "2024"
}

@article{punturo2010,
  title={The Einstein Telescope: a third-generation gravitational wave observatory},
  author={Punturo, M and Abernathy, M and Acernese, F and Allen, B and Andersson, Nils and Arun, K and Barone, F and Barr, B and Barsuglia, M and Beker, M and others},
  journal={Classical and Quantum Gravity},
  volume={27},
  number={19},
  pages={194002},
  year={2010},
  publisher={IOP Publishing}
}

@article{Reitze:2019iox,
    author = "Reitze, David and others",
    title = "{Cosmic Explorer: The U.S. Contribution to Gravitational-Wave Astronomy beyond LIGO}",
    eprint = "1907.04833",
    archivePrefix = "arXiv",
    primaryClass = "astro-ph.IM",
    reportNumber = "LIGO-P1900316",
    journal = "Bull. Am. Astron. Soc.",
    volume = "51",
    number = "7",
    pages = "035",
    year = "2019"
}

@article{Buonanno:2009zt,
    author = "Buonanno, Alessandra and Iyer, Bala and Ochsner, Evan and Pan, Yi and Sathyaprakash, B. S.",
    title = "{Comparison of post-Newtonian templates for compact binary inspiral signals in gravitational-wave detectors}",
    eprint = "0907.0700",
    archivePrefix = "arXiv",
    primaryClass = "gr-qc",
    doi = "10.1103/PhysRevD.80.084043",
    journal = "Phys. Rev. D",
    volume = "80",
    pages = "084043",
    year = "2009"
}

@article{Moore:2016qxz,
    author = "Moore, Blake and Favata, Marc and Arun, K. G. and Mishra, Chandra Kant",
    title = "{Gravitational-wave phasing for low-eccentricity inspiralling compact binaries to 3PN order}",
    eprint = "1605.00304",
    archivePrefix = "arXiv",
    primaryClass = "gr-qc",
    reportNumber = "LIGO-DCC-P1500268",
    doi = "10.1103/PhysRevD.93.124061",
    journal = "Phys. Rev. D",
    volume = "93",
    number = "12",
    pages = "124061",
    year = "2016"
}

@article{Khan:2018fmp,
    author = "Khan, Sebastian and Chatziioannou, Katerina and Hannam, Mark and Ohme, Frank",
    title = "{Phenomenological model for the gravitational-wave signal from precessing binary black holes with two-spin effects}",
    eprint = "1809.10113",
    archivePrefix = "arXiv",
    primaryClass = "gr-qc",
    doi = "10.1103/PhysRevD.100.024059",
    journal = "Phys. Rev. D",
    volume = "100",
    number = "2",
    pages = "024059",
    year = "2019"
}

@article{vanderWalt:2011bqk,
    author = "van der Walt, Stéfan and Colbert, S. Chris and Varoquaux, Gaël",
    archivePrefix = "arXiv",
    doi = "10.1109/MCSE.2011.37",
    eprint = "1102.1523",
    journal = "Comput. Sci. Eng.",
    number = "2",
    pages = "22--30",
    primaryClass = "cs.MS",
    title = "{The NumPy Array: A Structure for Efficient Numerical Computation}",
    volume = "13",
    year = "2011"
}

@inproceedings{jupyter,
       booktitle = {Positioning and Power in Academic Publishing: Players, Agents and Agendas},
          editor = {Fernando Loizides and Birgit Scmidt},
           title = {Jupyter Notebooks - a publishing format for reproducible computational workflows},
          author = {Thomas Kluyver and Benjamin Ragan-Kelley and Fernando P{\'e}rez and Brian Granger and Matthias Bussonnier and Jonathan Frederic and Kyle Kelley and Jessica Hamrick and Jason Grout and Sylvain Corlay and Paul Ivanov and Dami{\'a}n Avila and Safia Abdalla and Carol Willing and  Jupyter development team},
       publisher = {IOS Press},
         address = {Netherlands},
            year = {2016},
           pages = {87--90},
             url = {https://eprints.soton.ac.uk/403913/}
}

@article{Virtanen:2019joe,
    author = "Virtanen, Pauli and others",
    archivePrefix = "arXiv",
    doi = "10.1038/s41592-019-0686-2",
    eprint = "1907.10121",
    journal = "Nature Meth.",
    primaryClass = "cs.MS",
    title = "{SciPy 1.0--Fundamental Algorithms for Scientific Computing in Python}",
    year = "2020"
}

@Article{Hunter:2007,
  Author    = {Hunter, J. D.},
  Title     = {Matplotlib: A 2D graphics environment},
  Journal   = {Computing in Science \& Engineering},
  Volume    = {9},
  Number    = {3},
  Pages     = {90--95},
  publisher = {IEEE COMPUTER SOC},
  doi       = {10.1109/MCSE.2007.55},
  year      = 2007
}

@article{2013A&A...558A..33A,
       title={Astropy: A community Python package for astronomy},
   volume={558},
   ISSN={1432-0746},
   url={http://dx.doi.org/10.1051/0004-6361/201322068},
   DOI={10.1051/0004-6361/201322068},
   journal={Astronomy \& amp; Astrophysics},
   publisher={EDP Sciences},
   author={Robitaille, Thomas P. and others},
   year={2013},
   month=sep, pages={A33} 
}

@article{2018AJ....156..123A,
       title={The Astropy Project: Building an Open-science Project and Status of the v2.0 Core Package*},
   volume={156},
   ISSN={1538-3881},
   url={http://dx.doi.org/10.3847/1538-3881/aabc4f},
   DOI={10.3847/1538-3881/aabc4f},
   number={3},
   journal={The Astronomical Journal},
   publisher={American Astronomical Society},
   author={Price-Whelan, A. M. and others},
   year={2018},
   month=aug, pages={123} }

@software{pycbc,
  author       = {Alex Nitz and
                  Ian Harry and
                  Duncan Brown and
                  Christopher M. Biwer and
                  Josh Willis and
                  Tito Dal Canton and
                  Collin Capano and
                  Thomas Dent and
                  Larne Pekowsky and
                  Andrew R. Williamson and
                  Soumi De and
                  Miriam Cabero and
                  Bernd Machenschalk and
                  Duncan Macleod and
                  Prayush Kumar and
                  Steven Reyes and
                  Francesco Pannarale and
                  Gareth S Cabourn Davies and
                  dfinstad and
                  Sumit Kumar and
                  Márton Tápai and
                  Leo Singer and
                  Sebastian Khan and
                  Stephen Fairhurst and
                  Alex Nielsen and
                  Shashwat Singh and
                  Thomas Massinger and
                  Koustav Chandra and
                  Shasvath and
                  Veronica-villa},
  title        = {gwastro/pycbc: v2.0.4 release of PyCBC},
  month        = jun,
  year         = 2022,
  publisher    = {Zenodo},
  version      = {v2.0.4},
  doi          = {10.5281/zenodo.6646669},
  url          = {https://doi.org/10.5281/zenodo.6646669}
}

@article{KAGRA:2020tym,
    author = "Akutsu, T. and others",
    collaboration = "KAGRA",
    title = "{Overview of KAGRA: Detector design and construction history}",
    eprint = "2005.05574",
    archivePrefix = "arXiv",
    primaryClass = "physics.ins-det",
    doi = "10.1093/ptep/ptaa125",
    journal = "PTEP",
    volume = "2021",
    number = "5",
    pages = "05A101",
    year = "2021"
}

@article{LIGOScientific:2018mvr,
    author = "Abbott, B. P. and others",
    collaboration = "LIGO Scientific, Virgo",
    title = "{GWTC-1: A Gravitational-Wave Transient Catalog of Compact Binary Mergers Observed by LIGO and Virgo during the First and Second Observing Runs}",
    eprint = "1811.12907",
    archivePrefix = "arXiv",
    primaryClass = "astro-ph.HE",
    reportNumber = "LIGO-P1800307",
    doi = "10.1103/PhysRevX.9.031040",
    journal = "Phys. Rev. X",
    volume = "9",
    number = "3",
    pages = "031040",
    year = "2019"
}

@article{LIGOScientific:2020ibl,
    author = "Abbott, R. and others",
    collaboration = "LIGO Scientific, Virgo",
    title = "{GWTC-2: Compact Binary Coalescences Observed by LIGO and Virgo During the First Half of the Third Observing Run}",
    eprint = "2010.14527",
    archivePrefix = "arXiv",
    primaryClass = "gr-qc",
    reportNumber = "P2000061",
    doi = "10.1103/PhysRevX.11.021053",
    journal = "Phys. Rev. X",
    volume = "11",
    pages = "021053",
    year = "2021"
}

@article{LIGOScientific:2021usb,
    author = "Abbott, R. and others",
    collaboration = "LIGO Scientific, VIRGO",
    title = "{GWTC-2.1: Deep extended catalog of compact binary coalescences observed by LIGO and Virgo during the first half of the third observing run}",
    eprint = "2108.01045",
    archivePrefix = "arXiv",
    primaryClass = "gr-qc",
    reportNumber = "LIGO-P2100063",
    doi = "10.1103/PhysRevD.109.022001",
    journal = "Phys. Rev. D",
    volume = "109",
    number = "2",
    pages = "022001",
    year = "2024"
}

@article{KAGRA:2021vkt,
    author = "Abbott, R. and others",
    collaboration = "KAGRA, VIRGO, LIGO Scientific",
    title = "{GWTC-3: Compact Binary Coalescences Observed by LIGO and Virgo during the Second Part of the Third Observing Run}",
    eprint = "2111.03606",
    archivePrefix = "arXiv",
    primaryClass = "gr-qc",
    reportNumber = "LIGO-P2000318",
    doi = "10.1103/PhysRevX.13.041039",
    journal = "Phys. Rev. X",
    volume = "13",
    number = "4",
    pages = "041039",
    year = "2023"
}

@article{Chan:2024qmb,
    author = "Chan, Juno C. L. and Seo, Eungwang and Li, Alvin K. Y. and Fong, Heather and Ezquiaga, Jose M.",
    title = "{Detectability of lensed gravitational waves in matched-filtering searches}",
    eprint = "2411.13058",
    archivePrefix = "arXiv",
    primaryClass = "gr-qc",
    doi = "10.1103/PhysRevD.111.084019",
    journal = "Phys. Rev. D",
    volume = "111",
    number = "8",
    pages = "084019",
    year = "2025"
}

@article{Chakraborty:2024net,
    author = "Chakraborty, Aniruddha and Mukherjee, Suvodip",
    title = "{GLANCE \textendash{} Gravitational Lensing Authenticator using Non-modelled Cross-correlation Exploration of Gravitational Wave Signals}",
    eprint = "2403.03982",
    archivePrefix = "arXiv",
    primaryClass = "gr-qc",
    doi = "10.1093/mnras/stae1800",
    journal = "Mon. Not. Roy. Astron. Soc.",
    volume = "532",
    number = "4",
    pages = "4842--4863",
    year = "2024"
}

@article{Chakraborty:2024mbr,
    author = "Chakraborty, Aniruddha and Mukherjee, Suvodip",
    title = "{$\mu$-GLANCE: A Novel Technique to Detect Chromatically and Achromatically Lensed Gravitational Wave Signals}",
    eprint = "2410.06995",
    archivePrefix = "arXiv",
    primaryClass = "gr-qc",
    month = "10",
    year = "2024"
}

@article{Chakraborty:2025maj,
    author = "Chakraborty, Aniruddha and Mukherjee, Suvodip",
    title = "{The First Model-Independent Chromatic Microlensing Search: No Evidence in the Gravitational Wave Catalog of LIGO-Virgo-KAGRA}",
    eprint = "2503.16281",
    archivePrefix = "arXiv",
    primaryClass = "gr-qc",
    month = "3",
    year = "2025"
}

\end{document}